\documentclass[acmtog]{acmart}
\acmSubmissionID{169}

\usepackage{booktabs} 
\usepackage{color, colortbl}
\definecolor{lightgray}{gray}{0.9}
\usepackage{xcolor}
\usepackage{wrapfig}
\usepackage{enumitem}
\usepackage{bm}

\usepackage{amsmath}
\makeatletter
\let\save@mathaccent\mathaccent
\newcommand*\if@single[3]{%
  \setbox0\hbox{${\mathaccent"0362{#1}}^H$}%
  \setbox2\hbox{${\mathaccent"0362{\kern0pt#1}}^H$}%
  \ifdim\ht0=\ht2 #3\else #2\fi
  }
\newcommand*\rel@kern[1]{\kern#1\dimexpr\macc@kerna}
\newcommand*\widebar[1]{\@ifnextchar^{{\wide@bar{#1}{0}}}{\wide@bar{#1}{1}}}
\newcommand*\wide@bar[2]{\if@single{#1}{\wide@bar@{#1}{#2}{1}}{\wide@bar@{#1}{#2}{2}}}
\newcommand*\wide@bar@[3]{%
  \begingroup
  \def\mathaccent##1##2{%
    \let\mathaccent\save@mathaccent
    \if#32 \let\macc@nucleus\first@char \fi
    \setbox\z@\hbox{$\macc@style{\macc@nucleus}_{}$}%
    \setbox\tw@\hbox{$\macc@style{\macc@nucleus}{}_{}$}%
    \dimen@\wd\tw@
    \advance\dimen@-\wd\z@
    \divide\dimen@ 3
    \@tempdima\wd\tw@
    \advance\@tempdima-\scriptspace
    \divide\@tempdima 10
    \advance\dimen@-\@tempdima
    \ifdim\dimen@>\z@ \dimen@0pt\fi
    \rel@kern{0.6}\kern-\dimen@
    \if#31
      \overline{\rel@kern{-0.6}\kern\dimen@\macc@nucleus\rel@kern{0.4}\kern\dimen@}%
      \advance\dimen@0.4\dimexpr\macc@kerna
      \let\final@kern#2%
      \ifdim\dimen@<\z@ \let\final@kern1\fi
      \if\final@kern1 \kern-\dimen@\fi
    \else
      \overline{\rel@kern{-0.6}\kern\dimen@#1}%
    \fi
  }%
  \macc@depth\@ne
  \let\math@bgroup\@empty \let\math@egroup\macc@set@skewchar
  \mathsurround\z@ \frozen@everymath{\mathgroup\macc@group\relax}%
  \macc@set@skewchar\relax
  \let\mathaccentV\macc@nested@a
  \if#31
    \macc@nested@a\relax111{#1}%
  \else
    \def\gobble@till@marker##1\endmarker{}%
    \futurelet\first@char\gobble@till@marker#1\endmarker
    \ifcat\noexpand\first@char A\else
      \def\first@char{}%
    \fi
    \macc@nested@a\relax111{\first@char}%
  \fi
  \endgroup
}
\usepackage[mathscr]{euscript}

\usepackage[ruled]{algorithm2e} 

\SetAlFnt{\small}
\SetAlCapFnt{\small}
\SetAlCapNameFnt{\small}
\SetAlCapHSkip{0pt}

\newcommand{\red}[1]{\textcolor{red}{#1}}

\newcommand{\blue}[1]{\textcolor{blue}{#1}}

\newcommand{\teal}[1]{\textcolor{teal}{#1}}

\newcommand{\R}{\mathbb{R}} 

\newlength\savedwidth
\newcommand\whline[1]{\noalign{\global\savedwidth\arrayrulewidth
                               \global\arrayrulewidth #1} %
                      \hline
                      \noalign{\global\arrayrulewidth\savedwidth}}
\acmJournal{TOG}

\begin{document}

\title{Affine Body Dynamics: \\Fast, Stable and Intersection-free Simulation of Stiff Materials}

\author{Lei Lan}
\affiliation{\institution{Clemson University}}
\email{lan6@clemson.edu}

\author{Danny M. Kaufman}
\affiliation{\institution{Adobe Research}}
\email{dannykaufman@gmail.com}

\author{Minchen Li}
\affiliation{\institution{UCLA}}
\email{minchernl@gmail.com}

\author{Chenfanfu Jiang}
\affiliation{\institution{UCLA}}
\email{chenfanfu.Jiang@gmail.com}

\author{Yin Yang}
\affiliation{\institution{Clemson University}}
\email{yin5@clemson.edu}

\begin{abstract}
Simulating stiff materials in applications where deformations are either not significant or else can safely be ignored is a fundamental task across fields. Rigid body modeling has thus long remained a critical tool and is, by far, the most popular simulation strategy currently employed for modeling stiff solids. At the same time, rigid body methods continue to pose a number of well known challenges and trade-offs including intersections, instabilities, inaccuracies, and/or slow performances that grow with contact-problem complexity. In this paper we revisit the stiff body problem and present ABD, a simple and highly effective affine body dynamics framework, which significantly improves state-of-the-art for simulating stiff-body dynamics. We trace the challenges in the rigid-body methods to the necessity of linearizing piecewise-rigid trajectories and subsequent constraints. ABD instead relaxes the unnecessary (and unrealistic) constraint that each body's motion be exactly rigid with a stiff orthogonality potential, while preserving the rigid body model's key feature of a small coordinate representation. In doing so ABD replaces piecewise \emph{linearization} with piecewise \emph{linear} trajectories. This, in turn, combines the best of both worlds: compact coordinates ensure small, sparse system solves, while piecewise-linear trajectories enable efficient and accurate constraint (contact and joint) evaluations. Beginning with this simple foundation, ABD preserves all guarantees of the underlying IPC model we build it upon, e.g., solution convergence, guaranteed non-intersection, and accurate frictional contact. Over a wide range and scale of simulation problems we demonstrate that ABD brings orders of magnitude performance gains (two- to three-order on the CPU and an order more utilizing the GPU, obtaining $10,000\times$ speedups) over prior IPC-based methods with a while maintaining simulation quality and nonintersection of trajectories. At the same time ABD has comparable or faster timings when compared to state-of-the-art rigid body libraries optimized for performance without guarantees, and successfully and efficiently solves challenging simulation problems where both classes of prior rigid body simulation methods fail altogether.

\end{abstract}



\newcommand{\gp}[1]{{\left({#1}\right)}}
\newcommand{\ab}[1]{{\left|{#1}\right|}}
\newcommand{\bk}[1]{{\left[{#1}\right]}}
\newcommand{\px}[2]{\frac{\partial #1}{\partial #2}}
\newcommand{\pxx}[2]{\frac{\partial^2 #1}{\partial {#2}^2}}
\newcommand{\pxy}[3]{\frac{\partial^2 #1}{\partial #2 \partial #3}}
\newcommand{\mx}[1]{\begin{pmatrix}#1\end{pmatrix}}
\newcommand{\xx}{\mathbf{x}}
\renewcommand{\tt}{\mathbf{t}}
\newcommand{\yy}{\mathbf{y}}
\newcommand{\zz}{\mathbf{z}}
\newcommand{\uu}{\mathbf{u}}
\newcommand{\ff}{\mathbf{f}}
\newcommand{\ww}{\mathbf{w}}
\newcommand{\FF}{\mathbf{F}}
\newcommand{\XX}{\mathbf{X}}
\renewcommand{\AA}{\mathbf{A}}
\newcommand{\BB}{\mathbf{B}}
\newcommand{\RR}{\mathbf{R}}
\newcommand{\DD}{\mathbf{D}}
\newcommand{\pp}{\mathbf{p}}
\newcommand{\ZZ}{\mathbf{Z}}
\newcommand{\rr}{\mathbf{r}}
\newcommand{\CC}{\mathbf{C}}
\newcommand{\UU}{\mathbf{U}}
\newcommand{\VV}{\mathbf{V}}
\newcommand{\cc}{\mathbf{c}}
\newcommand{\II}{\mathbf{I}}
\newcommand{\nn}{\mathbf{n}}
\newcommand{\TT}{\mathbf{T}}
\renewcommand{\ss}{\mathbf{s}}
\newcommand{\NN}{\mathbf{N}}
\newcommand{\EE}{\mathbf{E}}
\newcommand{\HH}{\mathbf{H}}
\newcommand{\PP}{\mathbf{P}}
\newcommand{\QQ}{\mathbf{Q}}
\newcommand{\MM}{\mathbf{M}}
\newcommand{\qq}{\mathbf{q}}
\renewcommand{\gg}{\mathbf{g}}
\newcommand{\GG}{\mathbf{G}}
\newcommand{\dd}{\mathbf{d}}
\newcommand{\PhiPhi}{\mathbf{\Phi}}
\newcommand{\Th}{\hat{\TT}}
\renewcommand{\SS}{\mathbf{S}}
\newcommand{\Sig}{\boldsymbol{\Sigma}}
\newcommand{\sig}{\boldsymbol{\sigma}}
\newcommand{\Lam}{\boldsymbol{\Lambda}}
\newcommand{\ps}{\boldsymbol{\psi}}
\newcommand{\td}[1]{\textcolor{red}{\textbf{Todo: #1}}}
\newcommand{\cg}[1]{\textcolor{red}{\textbf{Change: #1}}}
\newcommand{\nt}[1]{\textcolor{red}{\textbf{Note: #1}}}

\definecolor{yellow-green}{rgb}{0.6, 0.8, 0.2}
\newcommand{\fanfu}[1]{\textcolor{yellow-green}{\textbf{Fanfu: #1}}}
\newcommand{\minchen}[1]{\textcolor{cyan}{\textbf{Minchen: #1}}}

\begin{teaserfigure}
  \center
  \includegraphics[width = \textwidth]{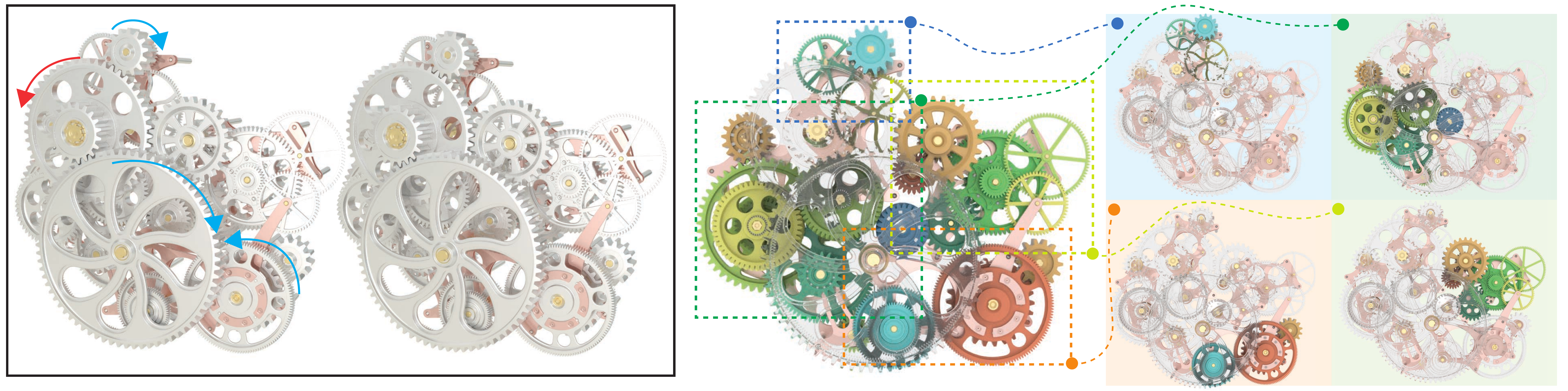}
  \caption{\textbf{Geared system.}~~We propose an affine body dynamics framework (ABD) to efficiently and robustly simulate close-to-rigid contacting objects. ABD significantly eases the collision and contact processing costs over rigid body modeling while obtaining intersection-free trajectories leveraging barrier-based frictional contact model. Here, in a challenging stress-test benchmark we simulate a complex geared system composed of 28 toothed gears with frictional contact resolving all interactions. The combined gear set mesh is comprised of over $2.45$M surface triangles. A torque applied to the actuated gear (highlighted in red arrow) drives the motion of the entire system via contact, with well over a quarter of all surface elements in active contact in every time step. Here we find that \emph{all the existing rigid body simulation algorithms, including rigid-IPC, fail to make progress}, while ABD robustly simulates the example to completion. ABD enables large time step sizes (e.g., with $\Delta t = 1/50$~sec) even for such challenging, large displacement (rotation) contact processing. For a time step size of $1/100$~sec, ABD simulates each step in less than $12$~sec on an \texttt{intel~i9} CPU (multi-threaded), while the simulation runs at an interactive rate on a \texttt{3090} GPU ($5$-$10$ time steps per second).}
  \label{fig:teaser}
\end{teaserfigure}

\maketitle
\section{Introduction}\label{sec:intro}

The simulation of highly stiff and so close-to-rigid materials remains a critical task in diverse applications ranging from animation and computer vision to robotics and geomechanics. A long-standing and natural strategy then is to model these bodies at the limit of stiffness and so treat them as exactly rigid. Equipped with just rotational and translational degrees of freedom (DOFs) this simplification enables the computational efficiency of classic rigid body methods as they utilize orders of magnitude fewer DOFs when deformations can be excluded.

At the same time this minimal representation for rigid bodies poses several fundamental difficulties of its own in exchange for optimized system size. The first being that moving from a flat continuum model to rigid (SE(3))
coordinates introduces significant nonlinearities that must be resolved. The second being that infinite stiffness implies applied forces, and especially contact responses, are communicated instantaneously across the material domain. In combination these two issues have long challenged both rigid body models (e.g., Painlev\'e's paradox) and the downstream simulation methods derived from them. State-of-the-art rigid body methods have long focused on efficient velocity (twist) level solutions. However, in doing so, these methods tend to generate undesired positional errors in the form of intersections and instabilities.

To address these issues a \emph{position-level} barrier method for rigid bodies was recently introduced by Ferguson and colleagues~\shortcite{ferguson2021intersection}. Applying the incremental potential contact (IPC) model~\cite{li2020incremental} to the rigid body formulation, the resulting rigid-IPC method provides robust intersection-free simulation of rigid solids with frictional contact. However, here a third challenge posed by rigid body models limits the efficiency and applicability of rigid-IPC: accurately tracing a piecewise-rigid trajectory is much more difficult than for a piecewise-linear trajectory. Unfortunately, doing so across numerous continuous-collision detection (CCD) operations throughout computation is a must for the non-intersecting guarantee. Rigid-IPC addresses this third challenge by conservatively subdividing rigid transformations into piecewise-linear subsequences i.e., the curved CCD. While proven to be viable, the expense remains substantial -- especially as CCD is invoked heavily throughout each simulation step. As a result, the overall performance of rigid-IPC is close to (and occasionally slower than) comparable (appropriately stiffened) fullspace finite element method (FEM) IPC simulations~\cite{li2020incremental}. Here the curved CCD in rigid-IPC severely undermines the advantage of its small DOF representation.

Beginning with this analysis, our takeaways are that the key advantage of rigid body models is \emph{not} the rigidity assumption but rather the compact representation. Indeed, the rigidity assumption is neither necessary, efficient, nor particularly accurate since no material is perfectly rigid. Following this reasoning, we approach the classic stiff-body problem from a different perspective, without the dedicated constraint and resultant limitations that the motion be exactly rigid. Specifically, we construct an affine-body dynamics  (ABD) model, directly stiffened to obtain close-to-rigid trajectories and augmented with an IPC-type barrier. ABD preserves all guarantees of the IPC model including solution convergence, guaranteed non-intersection, and accurate frictional contact. However, discrete steps in ABD are now, as in the FEM case, piecewise linear, enabling us to utilize efficient linear CCD routines. Likewise ABD remains compact, utilizing 12 DOFs per body -- a bit more than rigid bodies but still compact enough for efficient system solves. In turn the relaxation from the rigidity constraint allows ABD to significantly outperform the rigid-IPC method across all benchmarks (ranging from two- to three-order speedups for side-by-side comparison on the CPU, and an order-of-magnitude further improvement enabled by our GPU implementation), and to likewise successfully simulate problems where rigid-IPC fails.

At the same time, by combining compact representation and efficient collision processing, ABD also exhibits clear advantages in quality, reliability and even performance when compared to off-the-shelf rigid body simulation libraries. Here these libraries (we use \texttt{Bullet}~\cite{coumans2015bullet} as our baseline for comparison; see Section\ \ref{sec:result} for a discussion of this choice) are often optimized for speed over robustness and guarantees. Nevertheless, without requiring the precomputation of collision proxies (e.g., the convex decompositions required by \texttt{Bullet}, \texttt{Mujoco}~\cite{todorov2012mujoco}, \texttt{PhysX} and \texttt{Flex}~\cite{nvidia2011physx}), ABD remains closely competitive in performance on small-scale examples, while obtaining significantly faster simulations on larger and/or more challenging scenarios.

ABD is also able to simulate a wide range of challenging modeling problems where existing rigid body methods and libraries fail altogether. As an example, in Fig.~\ref{fig:teaser} we demonstrate a driven mechanical system with all gear-to-gear interactions processed directly via frictional contact. As we apply an external torque to the driving gear, the entire mechanism moves with well over a quarter of the mesh's $2.45$M triangles actively in contact during each time step. Here we find rigid-IPC (curved CCD failures) and \texttt{Bullet} (severe intersections) are both unable to simulate this mechanism even as we adjust algorithm settings and time step sizes conservatively. ABD simulates the scene robustly without algorithm tuning. Under a time step of $\Delta t = 1/100$~sec, ABD simulates each step/frame in $12$~sec on the CPU (multi-threaded) and reaches an interactive speed on the GPU ranging from $5$ to $10$ FPS.

In a nutshell, ABD provides a new stiff-body simulation framework suitable for all rigid-body-type modeling problems that offers similar (or improved) performance when compared to existing rigid body libraries (optimized for performance), while providing guarantees of non-intersection, accurate contact, and convergent implicit solves that they do not. ABD does not require pre-computed convex part proxies for simulation geometries; frees users from time-consuming per-scene parameter sweeps to find parameters that work; and ensures successful simulation completion in challenging cases where prior methods fail altogether.

Our technical contribution is simple and highly effective -- a relaxation replacing strictly rigid bodies by stiff affine bodies. This ``trivial'' strategy leads to non-trivial improvements on a fundamental simulation problem in terms of both efficiency and robustness. Here, based our above analysis of prior methods, we see that our resulting ABD framework balances tradeoffs between compact representation, nonlinearity, rigidity, and reliable guaranteed contact resolution. 
In addition to this high-level idea, we also design a novel collision culling algorithm dedicated for our simulation and a corresponding parallel matrix assembly strategy. We demonstrate the efficacy and efficiency of ABD across extensive benchmark testing and comparisons. Keeping all favored guarantees, the observed speedup over the state-of-the-arts is often two- to three-order, i.e., over $1,000\times$ faster, and reaches $10,000\times$ on a GPU. We hope ABD will provide the rapidly growing and diverse community of simulation users with a reliable and efficient tool suitable to swap in for all rigid body-type applications.

\section{Related Work}\label{sec:related}
A rigid body is an idealization, which simplifies a highly stiff object. External stimulus like forces or impulses are propagated almost instantaneously across the object eliminating relative deformations. This property is normally handled \emph{kinematically} i.e., by directly formulating rigid body models with SE(3) coordinates. Rigid body models have been extensively studied by the graphics community dating back to the pioneering work of Baraff\ \shortcite{baraff1989analytical}. We refer the reader to a comprehensive survey from Bender and colleagues~\shortcite{bender2014interactive}, which covers a wide spectrum of classic rigid body simulation techniques.

The primary focus of rigid body problems is typically at the resolution of the collision and contact. Intuitively, objects should never intersect with each other at any instance during the simulation. Enforcing this requirement has traditionally lead to algorithms based on linear complementarity programming (LCP)~\cite{baraff1994fast,baraff1995interactive,trinkle2001dynamic,stewart2000rigid}. One needs to carefully search within the combinatorial space for an approximation of a feasible configuration, while the interaction among bodies in contact is generally based on impulses rather than forces~\cite{baraff1989analytical}. The LCP-based contact problem is known to be NP-Hard due to the indeterminacy of which contacting nodes are contributing the collision impulse~\cite{baraff1991coping}. Alternate solvers formulated on approximations of velocity-level LCPs have also been popular ~\cite{erleben2007velocity,kaufman2005fast,anitescu1997formulating} for improved solvability. Yet, the resulting system remains non-convex and challenging to solve with accuracy for complex scenes. Irrespective of the accuracy of the solution, here the required constraint linearization means that intersections leading downstream to artifacts like drifting and tunneling can and will result. To reduce these artifacts, additional constraint stabilizations are often employed~\cite{cline2003post,baumgarte1972stabilization,moreau1988unilateral} however this, in turn can introduce new instability artifacts like popping. In addition to rigid bodies, LCP modles are also widely used for contact modeling among non-rigid objects~\cite{hauser2003interactive,pauly2004quasi,song2003distributed,duriez2005realistic,otaduy2007adaptive}.

For small-scale problems, a direct LCP solver could be used. When the complexity and the dimensionality increase, iterative LCP solvers stand as a more efficient option. Here successful designs of various iterative methods for LCP-based contacts such as Gauss-Seidel~\cite{erleben2007velocity}, PROX (iterative proximal operator)~\cite{erleben2017rigid}, surrogate constraints~\cite{kaufman2005fast}, accelerated gradient descent~\cite{mazhar2015using}, staggered projection~\cite{kaufman2008staggered}, and adaptive merging~\cite{coevoet2020adaptive} have all been applied.

Penalties are also another popular option widely used to process collisions~\cite{terzopoulos1987elastically,teschner2005collision}. Instead of imposing inequality constraints, a penalty method often chooses a spring-like repulsion mechanism based on the penetration depth between two objects~\cite{fisher2001fast,hasegawa2004real,drumwright2007fast}. While computationally simple, the penalty method fails for fast-moving models or simulations under large time steps and often requires significant manual tuning of stiffness parameters per scene. Its stability can be significantly enhanced using implicit formulations coupled with CCD~\cite{xu2014implicit,tang2012continuous}. Nevertheless, interpenetration still can and will result. This defect limits its wider use beyond graphics, where visual plausibility is not the only concern. Recently, M\"uller and colleagues also proposed a position-based rigid body framework~\cite{muller2020detailed}. Unlike classic rigid body algorithms, this method uses PBD-like constraint projection~\cite{muller2007position,macklin2016xpbd} to process multiple-body dynamics.

Collision detection is another important procedure for modeling rigid bodies with contacts. In general, a collision could occur between any triangle pair of two objects, and an exhaustive triangle-based collision detection is infeasible for high-resolution models. To this end, a commonly adopted method is to use some bounding volume hierarchy (BVH)~\cite{zachmann2003geometric} to avoid excessive triangle-triangle intersection tests. This pre-screening procedure is known as collision culling. Different BV types have been explored such as AABB~\cite{bergen1997efficient}, OBB~\cite{gottschalk1996obbtree}, bounding sphere~\cite{hubbard1995collision,james2004bd}, Boxtree~\cite{zachmann2002minimal}, spherical shell~\cite{krishnan1998rapid} and so on. As the geometry of the model does not change in rigid body models, BVH updates become particularly convenient -- the per-body rigid transformation can be directly applied to update the BVH instead of re-building it from scratch (as opposite to deformable objects). In some existing rigid body packages e.g., \texttt{Bullet} library~\cite{coumans2015bullet}, the collision detection does not apply to the surface triangles directly but rather to a volumetric proxy of the model, formed of convex components -- most often obtained by via a convex decomposition. While methods like \texttt{Bullet} require these convex proxies for robust processing this proxy also helps as an acceleration for collision detection. In this paper, we provide a new culling method to accelerate for collision detection for rigid bodies and ABD. Our method leverages the fact that the colliding region between two rigid bodies often constitutes a very small fraction of their surfaces. Based on this observation, we create a BVH only covering the overlapping region of two bodies for a more effective culling.

Discrete collision detection (DCD) checks for collisions or penetrations at a specific time instance. This method could miss inter-penetrations if the detection is not performed frequently enough. Alternatively, CCD checks the possible overlap of the trajectories of the surface primitives and returns the first time of impact (TOI)~\cite{bridson2002robust,redon2005fast}. The overlap test for triangle-vertex and edge-edge becomes a cubic polynomial, and there are several root-finding algorithms are available for solving the TOI~\cite{redon2005fast,brochu2012efficient,tang2014fast}. In a recent contribution from Wang and colleagues~\shortcite{wang2021large}, a more stable root-finding algorithm was proposed based on an improved inclusion.

Lastly in this section, we would like to discuss our most closely related work on rigid-IPC from Ferguson and colleagues~\shortcite{ferguson2021intersection}. Rigid-IPC features a new rigid body formulation, where contact is modeled with a barrier-based potential i.e., the IPC~\cite{li2020incremental,li2020robust} model. Conceptually, IPC is similar to the implicit penalty method (e.g., as in~\cite{tang2012continuous}), which produces a repulsion force pushing apart two contacting objects. However, due to the dedicated design of the barrier function, IPC provides guaranteed intersection-free collision resolution when appropriately combined with a CCD-filtered linear search. When compared to existing contact handling methods, IPC has demonstrated a superior performance -- it is significantly faster than LCP-based solutions for complicated contacts where LCP-based methods often fail altogether and much more robust than regular penalty methods with user-specified accuracy bounds. This method has been broadly applied to elastodynamics simulation~\cite{li2020incremental}, codimensional models~\cite{li2020codimensional}, embedded interfaces~\cite{choo2021barrier}, FEM-MPM coupling~\cite{li2021bfemp}, deformation processing~\cite{Fang2021IDP}, and reduced models~\cite{lan2021medial}. Rigid-IPC transplants classic rigid body models to the framework of IPC providing significantly improved reliability in contact processing. Unfortunately, this also comes at a cost -- strictly rigidity motion imposes significant computational challenges. The trajectory in rigid-IPC becomes curved. In order to successfully compute TOI in curved CCD, Ferguson and colleagues~\shortcite{ferguson2021intersection} subdivide the rigid trajectory into piece-wise line segments, which becomes the new bottleneck of the simulation. We argue that the success of IPC has already proven the feasibility of using smooth approximation to substitute hard constraints (even the approximation itself could be stiff), why not use another smooth approximation for the rigidity constraint in traditional rigid body modeling? ABD is then devised following this intuition. The results are exciting: ABD is three-order faster than rigid-IPC. With our novel culling algorithm and parallel Hessian assembly, the speedup over rigid-IPC scales even further. Likewise, as ABD provides stiff compliance (as in real-world stiff materials) tight almost rigid parts robustly fit together where an absolutely strict rigidity model can fail. Finally, as in rigid-IPC, ABD provides user-controllable solution and contact accuracies with solutions always reaching the specified tolerances.

\section{ABD Kinematics}
\label{sec:affine}
We begin by constructing a flat kinematics via affine coordinates. We equip each simulated body $b$ in our domain with a time-varying linear transform $\mathsf{A}_b(t) \in \R^{3 \times 3}$, and a translation $p_b(t) \in \R^3$. In the following we often store per-body configuration in the vector form as: $
q = \left(p^T, a_1^T, a_2^T, a_3^T\right)^T \in \R^{12}$, with the transform $\mathsf{A} = [a_1, a_2, a_3]^T$, then stored in row order.

Each material point $k$ in body $b$ has a body frame (equivalently rest) position $\bar{x}_k$ with its corresponding world frame coordinates given by the affine map: 
\begin{equation}\label{eq:kinematic}
    x_k = \mathsf{A}_b \bar{x}_k + p_b = \mathsf{J}(\bar{x}_k) q, 
\end{equation}
and its velocity, 
\begin{equation}\label{eq:kinematic-vel}
    {\dot x}_k = \dot{\mathsf{A}}_b \bar{x}_k + \dot{p}_b = \mathsf{J}(\bar{x}_k) \dot q. 
\end{equation}
Here note that $\mathsf{J}(\bar{x}) = [ \mathsf{I}_3, \mathsf{I}_3 \otimes \bar{x}]$ is \emph{constant} across all configuration changes, where $\mathsf{I}_3$ is a 3 by 3 identity matrix.

\subsection{Kinetic Energy}
Given a mass density distribution, $\rho$, over the body domain, $\Omega$, the kinetic energy of each affine body is then
\begin{align}
\begin{split}
\frac{1}{2}   \int_\Omega  \rho \dot{x}^T \dot{x} \> \mathrm{d}\Omega &= \frac{1}{2}  \int_\Omega \rho (\dot{\mathsf{A}} \bar{x}  + \dot{p})^T (\dot{\mathsf{A}} \bar{x}  + {\dot p})\> \mathrm{d}\Omega \\
  &= \frac{1}{2}  \dot{q}^T \left( \int_\Omega \rho \> \mathsf{J}(\bar{x})^T \mathsf{J}(\bar{x}) \> \mathrm{d}\Omega \right) \dot{q}, 
  \end{split}
\end{align}
with the generalized mass matrix for each affine body defined as: 
\begin{align}
\label{eq:mass}
\mathsf{M} =  \int_\Omega \rho \> \mathsf{J}(\bar{x})^T \mathsf{J}(\bar{x}) \> \mathrm{d}\Omega.
\end{align} 
Here we note an additional convenience of ABD: for flat affine coordinates we obtain a constant mass matrix. In turn this ensures that the equations of motion for affine bodies, unlike rigid bodies, \emph{do not} add nonlinear Coriolis-type forces. Instead, they are embodied as generalized internal forces. 

With $V$ the total potential energy the free ABD is then simply the equations of motion: 
\begin{align}
\label{eq:EL}
\mathsf{M} \ddot{q}  =-\nabla V(q) + f,
\end{align}
where external forces $f_k \in \R^3$, applied at material points $k$, are included as $f = \sum_k \mathsf{J}(\bar{x}_k)^T f_k$.

\subsection{Orthogonality Potential}

In place of SE(3) coordinates we rigidify each affine body with a stiff orthogonality potential
\begin{equation}\label{eq:orthogonality}
V_{\perp}(q) = \kappa \nu \| \mathsf{A}\mathsf{A}^T - \mathsf{I}_3 \|^2_{F},    
\end{equation} 
scaled by the stiffness $\kappa$ and the body's volume $\nu$. We apply a large stiffness ($\kappa >100$GPa) to ensure the deformation on the body is sufficiently suppressed and negligible. 

As we relax the rigidity constraint, we instead apply a highly stiff penalty term. Generally engineering rule of thumb suggests that stiff penalties should be avoided in simulation, as they tend to exacerbate the numerical difficulties of computing with nonlinear potentials. However, we observe that contact and collision forces (irrespective of whether they are treated via constraints, penalties springs, or barriers) introduce a much more dominant stiffness to the system. In our case this already requires handling system solves with a robust Newton-type algorithm so that the overhead of a single, additional stiff orthogonality potential per body is indifferent. In practice, as we show in Section\ \ref{sec:result}, directly stiffened affine systems significantly improve in performance over comparable systems that resolve rigid motion explicitly.

The stiff potential itself provides the effective constitutive model for the affine bodies -- modulating their collision response upon impact (and so intrinsically handling restitution). While $V_{\perp}$ is a natural choice for close-to-rigid motion~\cite{moser1991discrete}, it is never exclusively suited. Indeed affine bodies could alternately be equipped with the ARAP energy~\cite{igarashi2005rigid,alexa2000rigid,sorkine2007rigid}, i.e., $\| \mathsf{A} - \mathsf{R}(\mathsf{A}) \|^2_F$ for $\mathsf{A} = \mathsf{R}\mathsf{S}$ being the polar decomposition, to enforce rigidity, or else neo-Hookean, or any number of other rotation-invariant hyperelastic energies~\cite{bonet1997nonlinear}. However, we find that the orthogonality potential is both effective and efficient. $V_{\perp}$ requires no expensive decompositions (as opposite to ARAP) -- it can be computed as a polynomial 
\begin{equation}\label{eq:ortho-energy}
V_{\perp} = \kappa \nu  \left(\sum ( a_i \cdot a_i - 1)^2 + \sum_{i \neq j} (a_i \cdot a_j)^2\right),
\end{equation}
leading to more efficient evaluations of energy gradient and Hessian:
\begin{align}\label{eq:ortho-poly}
\begin{split}
    \frac{\partial V_{\perp}}{\partial a_i} &= 2 \kappa \nu \left(2(a_i \cdot a_i - 1) a_i + 2\sum(a_j \otimes a_j)a_i\right), \\
    \frac{\partial^2 V_{\perp}}{\partial a_i^2} &= 2 \kappa \nu \left(4a_i \otimes a_i + 2(\|a_i\|^2-1) \mathsf{I}_3 + 2\sum a_j \otimes a_j \right).
    \end{split}
\end{align}
In comparison, energy operations for affine bodies with $V_{\perp}$ are over $43\%$ and $178\%$ faster than applying ARAP and neo-Hookean models respectively.

\section{Affine IPC}

We simulate systems of affine bodies with triangulated boundaries. Following Li and colleagues~\shortcite{li2020incremental}, for each affine body $b \in \mathscr{B}$, we construct a discrete incremental potential (IP), $E_b$, whose stationary points give the unconstrained time step update: 
\begin{equation}
q_b^{t+1} = \arg\min_{q_b} E_b(q_b), \quad E_b = \frac{1}{2} \|q_b - \widetilde{q}_b\|_{\mathsf{M}}^2+ \Delta t^2  V_{\perp}(q_b). 
\end{equation} 
Here, $\Delta t$ is the time step size. $\widetilde{q}_b = q_b^t + \Delta t \dot{q}_b^t + \Delta t^2 \mathsf{M}^{-1} f_b^{t+1}$ is a known vector depending on the body state from the previous step. $f_b$ is the per-body external force (i.e., in Eq.~\eqref{eq:EL}). Unlike rigid body dynamics\footnote{Rigid body models require Poisson or constrained Lagrangian methods for numerical time integration~\cite{hairer2006preprocessed}.}, flat equations of motion of ABD allows us to directly apply IPs for a broad range of standard implicit time integration methods~\cite{li2020incremental,li2020codimensional}.

Setting $q = \big(q_1^T,\cdots,q_{|\mathscr{B}|}^T\big)^T$ as the stacked vector of all the bodies, we construct IPC potentials for contact, $V_C(q)$,  and dissipative friction, $V_F(q)$, to model inter-body contact forces. The contact potential
\begin{align}
\label{eq:barrier}
V_C(q) = \kappa \sum_{i \in \mathscr{C}}  B\left(d_{i}(q)\right),
\end{align}
(stiffness $\kappa$) resolves contacts between all possible pairings $i \in \mathscr{C}$ of inter-body surface geometry primitives (e.g., edge-edge, vertex-face pairs between body meshes -- but no self-contact because of body's high stiffness). Here the smoothly clamped logarithmic barrier function
\begin{equation}
    B(d, \hat{d}) =
    \begin{cases}
        \displaystyle -(d - \hat{d})^2 \ln \left(\frac{d}{\hat{d}}\right), &0 < d < \hat{d} \\
        \displaystyle 0 & \quad \> \> \>  d \geq \hat{d}
    \end{cases},
    \label{eq:clamped-b}
\end{equation}
needs to evaluate the unsigned distances for all primitive pairs in $\mathscr{C}$. Smooth clamping ensures that proximities beyond $\hat{d}$ can be safely culled from evaluation without harming convergence, while surface pairs within the small, prescribed contact accuracy $\hat{d}$ are activated to receive contact forces. 

Likewise, the friction potential is defined as:
\begin{align}
\label{eq:barrier}
V_F(q) = \sum_{j \in \mathscr{F}} \mu \lambda_j m_j(q), 
\end{align}
where $\mathscr{F} \subseteq \mathscr{C}$ is the active subset of contact pairs with positive contact force magnitude $\lambda_j = - \kappa \nabla_q B\left(d_{j}(q)\right)$. $\mu$ is the coefficient of friction, and $m_j$ returns a mollified norm of the relative sliding velocity, orthogonal to the distance vector, between geometric pairs $j$. Li and colleagues~\shortcite{li2020incremental} demonstrated that $V_F$ provides smooth approximation of the nonsmooth Coulomb friction model with user-controlled accuracy. In turn, this obtains effective and accurate capture of frictional stick and slip behaviors for both maximal and reduced body models~\cite{li2020incremental,ferguson2021intersection}. 

At each time step we then construct a global IP for the full contact-coupled system as: $E = \sum E_b + V_C + V_F$, and solve for its minimizer as the updated system configuration.

\subsection{Affine CCD} 
To ensure that every search fulfils the nonintersection guarantee, a significant (and for rigid bodies dominant) cost of each time step's Newton solve is the repeated evaluation of CCD for primitive pairs in $\mathscr{C}$. Edge-edge or vertex-face pairs are defined on vertices $x_j \in \R^3$ for $j = 1$ to $4$ belonging to one of two bodies in $\mathscr{B}$. At any iteration $\ell$ of a Newton solve, new positions of these vertices $x_j$ are proposed, per participating body $b$, as $\mathsf{A}_b^{\ell} \bar{x}_j + p_b^\ell$, and so we search along the direction formed by $\Delta \mathsf{A}_b^{\ell} = \mathsf{A}_b^{\ell} - \mathsf{A}_b^{\ell - 1}$ and $\Delta p_b^{\ell} = p_b^{\ell} - p_b^{\ell-1}$. The corresponding trajectories are then just $(\mathsf{A}_b^{\ell - 1} + \alpha  \Delta \mathsf{A}_b^{\ell})  \bar{x}_j + p_b^{\ell} + \alpha \Delta p_b^{\ell - 1}$, with $\alpha \in [0,1]$. In other words, we are testing  displacements (from $0$ to $1$) along $\Delta \mathsf{A}_b^{\ell} \bar{x}_j + \Delta p_b^{\ell}$ starting from $x_{j}^{\ell-1} = \mathsf{A}_b^{\ell-1} \bar{x}_j + p_b^{\ell-1}$ i.e., the previous iteration footstep and so can apply standard linear CCD rather than the expensive, curved CCD required for rigid-body trajectories. Throughout we employ Additive CCD (ACCD) method~\cite{li2020codimensional} for all affine body CCD evaluations.

\subsection{Contact Culling via i-AABB}
\label{subsec:ibvh}
The IPC framework inverts traditional contact-processing practices by integrating collision detection within every iterate, inside each nonlinear time step solve (rather than once per time step, after completing a solve -- as has been a standard practice in rigid body pipelines). Above we have already seen the implications of this choice on CCD and the advantages for ABD. We next address culling of contact pair evaluations and then below, in the following section, an integrated approach to efficiently compute the local energy Hessian evaluations and their assembly. 

\begin{figure}[h!]
  \centering
  \includegraphics[width=\linewidth]{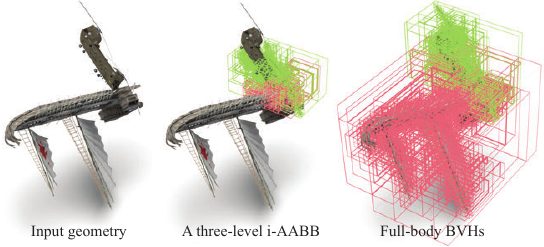}
  \caption{\textbf{i-AABB.}~~Intersection-AABB (i-AABB) is a simple and more effective culling strategy than per-object BVHs traditionally employed for rigid bodies. In this example, we collide helicopter and ship models with well over $343$K surface triangles for the system. Culling with full body BVHs leads to $1.3$M AABB intersection tests. Instead, we build a shallow, three-level i-AABB based on the models' overlapping AABB volume. With i-AABB, the total number of AABB intersection tests reduces to $120$K, $90\%$ fewer than regular BVH-based culling.}\label{fig:iBVH}
\end{figure}

With the affine bodies' stiff potentials applied to model close-to-rigid motion, we eliminate, as in rigid body models, the need for self-collision processing on intra-body surface primitives pairs. To further cull inter-body surface pairs from downstream collision processing, Rigid body methods commonly employ precomputed per-body (or per unique mesh instance) BVH which only requires a rigid (or, in our setting, affine) transform to evaluate at each configuration update. To further improve collision detection queries, most popular rigid body libraries (e.g., \texttt{Bullet}, \texttt{Mujoco}, \texttt{PhysX}, \texttt{Flex}) additionally require that all nonconvex surface meshes be replaced by approximations -- convex decompositions~\cite{mamou2016volumetric} so that evaluations can utilize convexity assumptions.

In the context of stiff-body models we propose a simple strategy that significantly enhances contact-pair culling. We start with the observation that contact-dense regions between close-to-rigid meshes are most often local. Unlike deforming meshes (e.g., consider a cloth model), in common, contact-rich configurations we see that a large portion of each body's surface remains free of contacts. A BVH evaluation over each body's full surface is then a bit too aggressive (and a potentially unnecessarily expensive precomputation cost) -- especially as we consider higher resolution models and/or increasing numbers of bodies. 

We instead begin with a coarse AABB per body, offset to account for the small $\hat{d}$ parameter of the contact accuracy. While clearly not providing a tight bound we only need to evaluate potential contacts in each pairwise overlap between these initial, \emph{intersecting} AABB (i-AABB) volume. With stiff affine transformations the overlap volume of each i-AABB is most often sufficiently small -- the total number of surface primitives within our i-AABB is often one order smaller than that obtained from full-mesh BVH queries. Then, utilizing the guarantee that i-AABBs will not intersect one another, this already provides effective culling for us to directly proceed with parallel (multi-threaded) primitive-pair collision-check evaluations within each i-AABB.

In the most extreme, heavily entangled configurations, e.g., see Fig.~\ref{fig:iBVH}, we find i-AABB volumes can be too conservative for CPU implementations. Here we simply build a shallow, three-level AABB hierarchy (a binary AABB tree) for each i-AABB. Our observation is that for GPU implementations and most (even contact-intense) CPU examples the single-level i-AABB is highly effective (and exceedingly simple to implement). We utilize the i-AABB hierarchies solely for CPU examples where dense, entangled contacts are consistently encountered, such as the geared system in Fig.~\ref{fig:teaser} and the interlocked collision of complex models in Fig.~\ref{fig:iBVH}. As a representative improvement we note that in the latter example the ship and the helicopter models have $227$K and $116$K surface triangles respectively; here the three-level hierarchical i-AABB is over $90\%$ more efficient in culling surface pairs in comparison to per-body BVH.

\subsection{Contact-Aware Hessian Construction}

It is, by far, most efficient to simulate a set of noncontacting affine bodies. Along with reduced cost for collision detection, the global (system-wide) Hessian for $E$ is then simply a  $12|\mathscr{B}| \times  12|\mathscr{B}|$ block-diagonal matrix with just a \emph{separable} $12 \times 12$ block for each affine body's mass and orthogonality energy ($V_{\perp}$) contributions. When the system includes contacts, however, off-diagonal terms from active, inter-body contact and friction potentials necessarily pollute the Hessian to account for contact coupling.

Standard FEM-type evaluation and assembly would suggest iterating across all active contact potentials. However, here the surface mesh resolution, and so the corresponding number of surface pairs forming contact potentials, is generally much larger than the number of affine bodies we simulate. Traditional assembly, in this contact-oblivious way, is neither necessary nor efficient. We instead integrate our Hessian evaluation and assembly with the i-AABB hierarchy for contact-aware parallelization in both multi-core CPU and GPGPU implementations.

We start with the easiest part. The global Hessian's default non-zero diagonal blocks are given by a constant mass term and the orthogonality potentials' Hessian (Eq.~\eqref{eq:ortho-poly}). This can be computed trivially in parallel. Next comes the expensive part. The Hessian of the contact potential, $V_C$, then has an unpredictable pattern which varies widely with contact states. Recall that, when bodies $i$ and $j$ are sufficiently close ($\leq \hat d$), barrier and friction forces between them activate, resulting in non-zero contributions to both their respective $i$-th and $j$-th $12 \times 12$ diagonal blocks \emph{and} to the off-diagonal blocks linking the corresponding body coordinates in the Hessian. 

Here we observe that the configuration-varying sparsity of the global Hessian is effectively determined by our i-AABB culling. If the leaves of the i-AABB tree between bodies $i$ and $j$ are empty, they certainly do not contact, and the barrier and friction potentials make no contribution to the Hessian. Otherwise, we can conservatively allocate space for the corresponding $12 \times 12$ off-diagonal blocks to store all potential (active contacts are not yet certain) non-zero contact Hessian contributions between bodies $i$ and $j$. Thus, utilizing the i-AABB structures we apply a two-pass strategy to compute and assemble barrier and friction terms for the global Hessian. The first pass iterates across all surface primitives pairs within each i-AABB; their local Hessians are computed in parallel and cached. Here we also account for the Hessian's symmetry to reduce total memory consumption. Our second pass then accumulates the local Hessians from the surface pairs. Here, as each culled i-AABB is independent and non-intersecting, accumulation is parallelized at each corresponding non-zero element of the global Hessian. 

\begin{figure*}[t!]
  \centering
  \includegraphics[width=\linewidth]{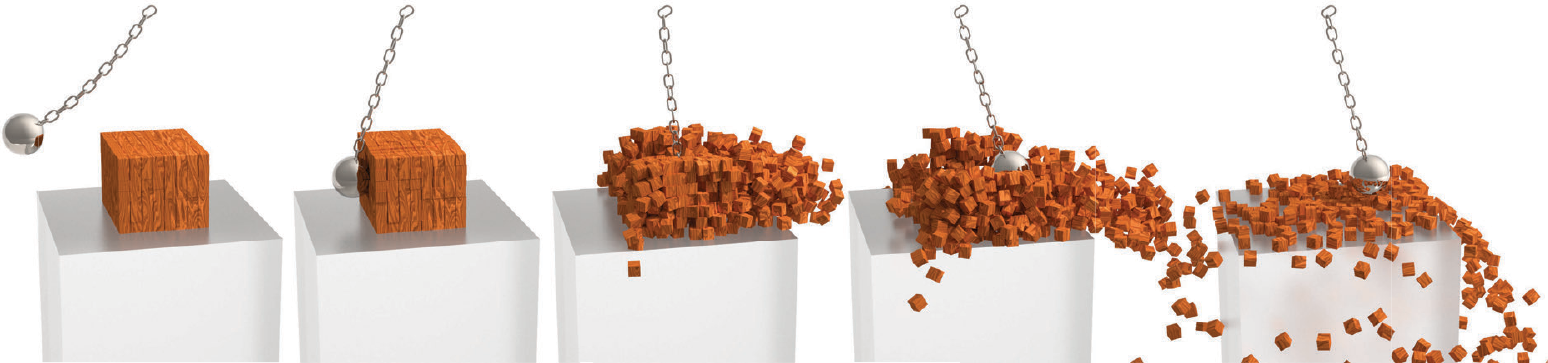}
  \caption{\textbf{Wrecking ball.}~~A metal ball linked to a chain of rigid rings hits a stack of $560$ wooden blocks. There are in total $575$ bodies including the ball, rings on the chain, and blocks in this example. ABD and rigid-IPC yield nearly identical simulation results but ABD is $124\times$ faster than rigid-IPC (both on CPU). The time step is $\Delta = 1/100$~sec. We also tested ABD using $\Delta = 1/50$~sec and $\Delta = 1/25$~sec respectively. The results are similar -- all are free of inter-penetration.}\label{fig:wrecking}
\end{figure*}

We find that i-AABB Hessian construction is significantly faster than default sequential parallelization of Hessian computation (e.g., as applied in rigid-IPC). We observe speedups of up to two orders, especially when the contacting system is composed of large numbers of bodies. Here, for example, the simulation in Fig.~\ref{fig:wrecking} provides a representative example, with a $188\times$ speedup over contact-oblivious, sequential Hessian construction.

\begin{figure*}[t!]
  \centering
  \includegraphics[width=\linewidth]{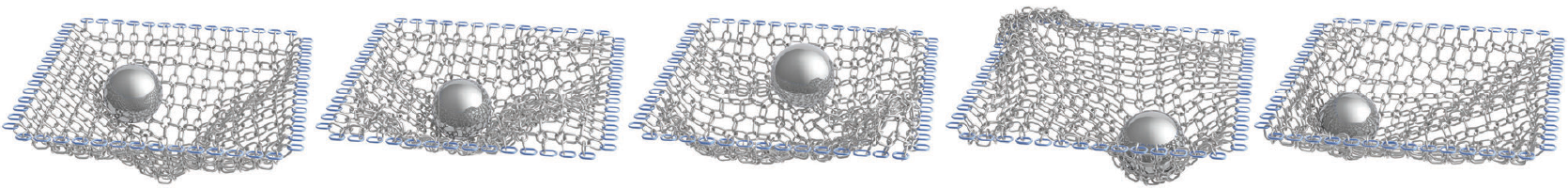}
  \caption{\textbf{Big chain net.}~~We increase the resolution of the chain net to $16 \times 16$ and drop a ball to the net. The ball hits the net and bounces back, which leads to interesting dynamical responses of the net. While both ABD and rigid-IPC are able to handle this simulation robustly, our method is $1,200\times$ faster on average. The speedup could reach over $10,000\times$ on GPU with \texttt{CUDA}.}\label{fig:net_big}
\end{figure*}

\section{Evaluation}\label{sec:result}
Our implementation platform is a desktop PC with an \texttt{intel i9 11900K} CPU (8 cores, 3.5GHZ), $64$G memory and an \texttt{nVidia~3090} GPU. All the numerical methods were implemented using \texttt{C++} on CPU, and we chose \texttt{Eigen}~\cite{guennebaud2010eigen} as our primary linear algebra library, including all sparse linear system solves. Our CPU parallelization utilizes \texttt{intel TBB}. We understand that some other BLAS libraries may be better optimized for our hardware (e.g., \texttt{intel MKL}~\cite{wang2014intel}). This choice is to ensure an objective comparison with rigid-IPC, whose multi-thread CPU implementation is also based on \texttt{Eigen} and \texttt{TBB}. We use Cholesky factorization~\cite{dereniowski2003cholesky} i.e., \texttt{SimplicialLLT} routine shipped with \texttt{Eigen} as our primary linear solver. In hybrid simulation problems, we may occasionally resort to \texttt{SimplicialLDLT} in case when LLT scheme fails (see more discussions in Section~\ref{subsec:hybrid}).  
\begin{figure}[ht!]
  \centering
  \includegraphics[width=\linewidth]{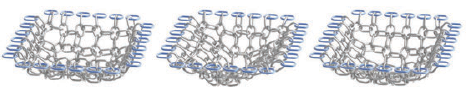}
  \caption{\textbf{Small chain net.}~~We simulate a small-size chain net consisting of $8$ by $8$ rings using both ABD and rigid-IPC. The exterior rings are bound with a loop of fixed blue rings. In this simple test, ABD is $34\times$ faster than rigid-IPC on the CPU. An increase of the time step size from $1/100$~sec to $1/50$~sec does not noticeably slow down our method (from $0.059$~sec to $0.065$~sec per frame), but rigid-IPC will be significantly slower (by $350\%$).}\label{fig:net}
\end{figure}

For benchmarking we focus on rigid-IPC~\cite{ferguson2021intersection} as comparison to the ``quality-oriented'' rigid body simulator with comparable guarantees to ABD and, as a representative baseline for comparison to state-of-the-art, optimized rigid body libraries we use \texttt{Bullet}~\cite{coumans2015bullet}. There are certainly numerous other, highly effective rigid body libraries with varying capabilities. However, trade-offs with respect to \texttt{Bullet} among alternatives have been extensively documented in~\cite{ferguson2021intersection}. In their comprehensive analysis and benchmark testing of rigid body libraries \texttt{Bullet} most consistently succeeds across challenging examples and, at the same time, we also note that \texttt{Bullet} is probably the most widely deployed rigid body solution.

\subsection{Comparison with Rigid-IPC}\label{subsec:comp_ripc}
\begin{figure}[h!]
  \centering
  \includegraphics[width=\linewidth]{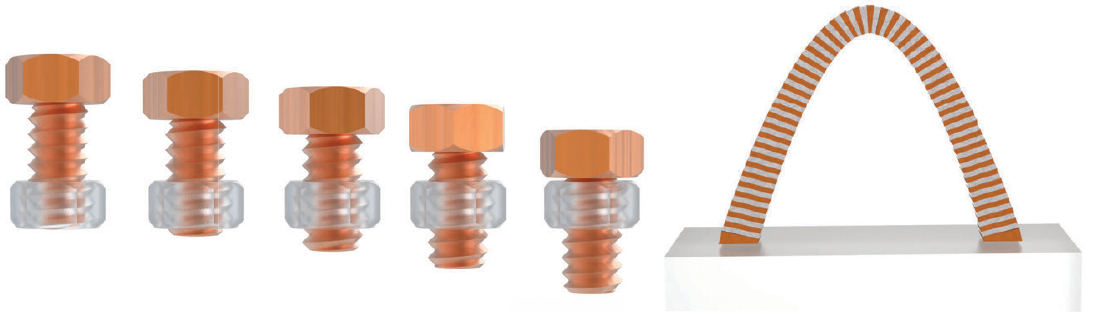}
  \caption{\textbf{Friction test.}~~While ABD and rigid-IPC use the same barrier-based friction processing method, ABD is much faster than rigid-IPC under the same simulation settings. The screw example (left) is a representative demonstration of dynamic friction/contact. Our speedup is $30\times$. On the other hand, the arch example (right) is dominated by static frictions, and ABD is $77\times$ faster than rigid-IPC.} \label{fig:friction}
\end{figure}

\begin{figure*}[t!]
  \centering
  \includegraphics[width=\linewidth]{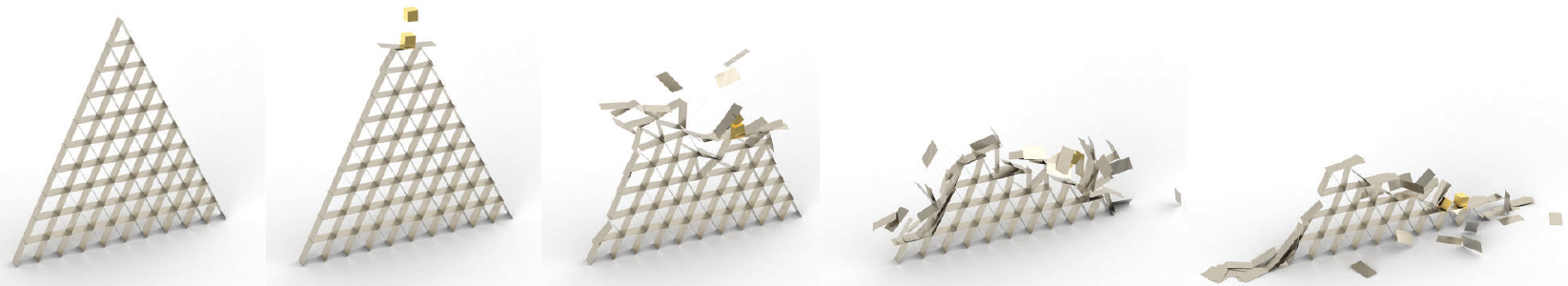}
  \caption{\textbf{House of cards.}~~The 10-level stack of cards is initially balanced by frictions among cards. Two falling boxes break the balance and crash the stack. This experiment mixes static and dynamic frictions involving $158$ bodies. Our method is $103\times$ faster than rigid-IPC.}\label{fig:cards}
\end{figure*}

Both ABD and rigid-IPC~\cite{ferguson2021intersection} exploit the IPC~\cite{li2020incremental} model for contact processing and friction modeling. Here we carefully compare ABD with rigid-IPC across several representative simulation scenarios of varying complexities, as illustrated in Figs.~\ref{fig:wrecking}~through~\ref{fig:cards}. 

The snapshots of the first comparison are given in Fig.~\ref{fig:wrecking}. In this test, a heavy ball linked by a chain of metal rings dashes into a stack of wooden blocks, which are scattered by the collision. There are in total $575$ bodies in this example. Both rigid-IPC and ABD produce high-quality simulation results without any inter-penetration. However, ABD is $124\times$ faster. The performance of ABD is not sensitive to the time step size. Doubling or even quadrupling the time step size (from $1/100$~sec to $1/50$~sec and $1/25$~sec) lead to similar results and performance with ABD. Detailed timing statistics can be found in Section~\ref{subsec:timing}. Similar observations are received in chain net examples. For the small chain net case (Fig.~\ref{fig:net}), we have few bodies, and the relative velocities among bodies are small. In this ``entry-level'' test, ABD offers a $34\times$ speedup. 

An ``upgraded'' experiment is reported in Fig.~\ref{fig:net_big}, where the resolution of the chain net is set to $16 \times 16$. We also drop a ball, which bounces back and forth on the net triggering interesting dynamics. The collisions and contacts also become more complicated than the plain simple chain net in Fig.~\ref{fig:net}. The difference between our method and rigid-IPC becomes more significant. At some instances when sharp collisions occur under high relative velocities, rigid-IPC could take multiple hours to simulate one frame ($\Delta t = 1/100$~sec). On the other hand, our method needs seconds at most. On average, ABD is over $1,200\times$ faster than rigid-IPC. This speedup exceeds $4,000\times$ from time to time during the simulation. In ABD, affine CCD, i-AABB, as well as the integrated Hessian assembly are all parallelization-ready. Therefore, ABD typically receives one more orders performance gain on \texttt{CUDA}. That makes ABD over $10,000\times$ faster than rigid-IPC in the example of Fig.~\ref{fig:net_big}.

The friction handling in ABD and rigid-IPC is similar as both follow the variational friction model originally proposed in~\cite{li2020incremental}. Nevertheless, ABD still exhibits superior performance in simulations dominated by frictions. We hereby report three more experiments in Figs.~\ref{fig:friction} and~\ref{fig:cards}. In those comparisons, both ABD and rigid-IPC use the same simulation settings with the time step size of $1/100$~sec. The screw example (Fig.~\ref{fig:friction} left) is relatively simple -- the surface geometry only has $7$K triangles and $5$K edges. We rotate the bolt into the nut. On average, rigid-IPC needs $2.59$~sec to simulate one step, and our method only needs $87$~ms. The arch (Fig.~\ref{fig:friction} right) consists of $100$ brick blocks. In this example, we mainly have static friction to deal with. Rigid-IPC runs faster than the screw case and needs $0.67$~sec for simulating one frame. Our method however, only uses $8.7$~ms. The card of house example starts with a 10-level stack of $155$ cards. The stack is first held up under the frictions between cards and gets crashed by two falling boxes. In this example, it takes about $8.9$~sec for rigid-IPC to simulate one frame, and ABD needs $86$~ms.

\begin{figure}[th!]
  \centering
  \includegraphics[width=\linewidth]{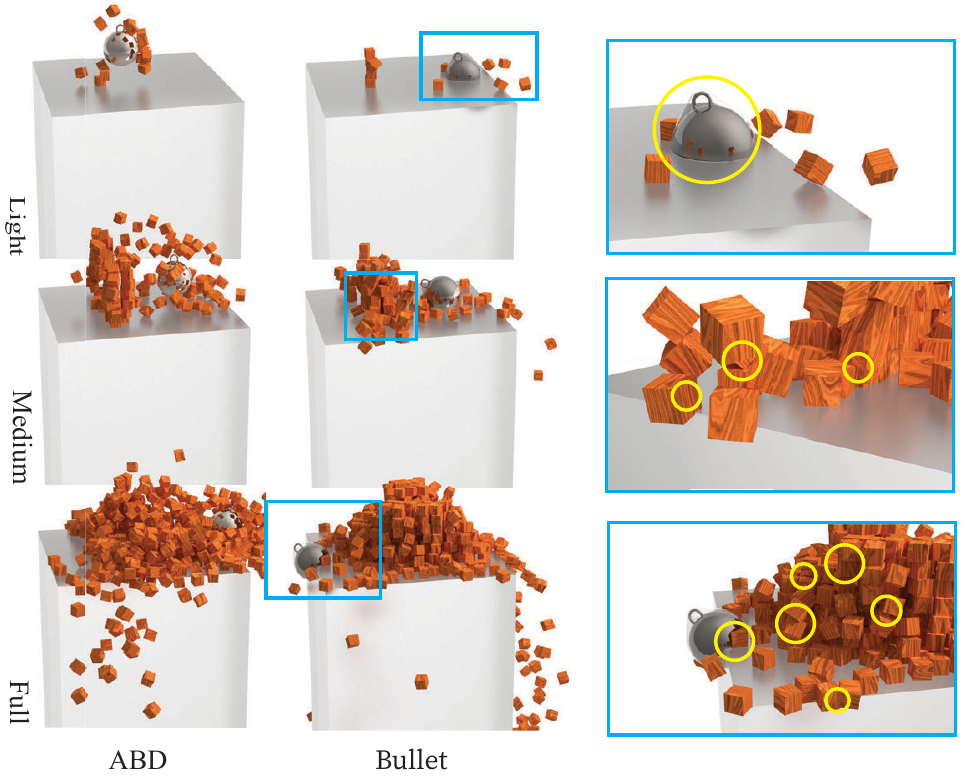}
  \caption{\textbf{ABD vs. \texttt{Bullet}.}~~We compare simulation results using ABD and \texttt{Bullet} with varying numbers of bodies and time step sizes. \texttt{Bullet} often produces interpenetrations between bodies even under small time steps. Interpenetrations becomes increasingly severe with growing body counts and/or time step size. On the other hand, ABD remains intersection-free across changing parameters and scene complexities. All timing statistics are reported in Tab.~\ref{tab:bullet}.} \label{fig:bullet}
\end{figure}

\subsection{Comparison with \texttt{Bullet}}
\texttt{Bullet} is a widely used rigid body engine known for the simplicity and efficiency. \texttt{Bullet} is also based on the classic rigid body model formulation. However, \texttt{Bullet} uses the so-called convex collision resolution, which approximates the geometry of the body via convex decomposition proxies. The collision is resolved using impulse-based method at the convex proxies. Therefore, it is not surprising to see \texttt{Bullet} fails in simulations under lasting and intense collisions and contacts. We have shown that our method is orders of magnitude faster than rigid-IPC with further improved robustness. For smaller simulation problems, ABD is nearly as efficient as \texttt{Bullet}. It quickly outperforms \texttt{Bullet} from almost all perspectives as the complexity of the simulation escalates.

We compare ABD and \texttt{Bullet} using the wrecking ball setup (i.e., see Fig.~\ref{fig:wrecking}) but under different block counts and time step sizes. In this set of comparisons, the ball is no longer attached to the chain as in Fig.~\ref{fig:wrecking}. This is because under high-velocity movements, collision resolution in \texttt{Bullet} frequently fails, and the rings on the chain would ``break out'' so that the ball may not hit the stack. The results are reported in Fig.~\ref{fig:bullet}, where we have three configurations: light, medium, and full. The light test only has $16$ blocks on the stack; the medium test has $142$ blocks; and the full test is the same as in Fig.~\ref{fig:wrecking} with $560$ blocks. 

\begin{table}[th!]
{\small \fontfamily{ppl}\selectfont
\begin{center}
\begin{tabular}{c|c|c|c|c|c}
\whline{1.15pt}
Test & $\#$ Bdy & $\#$ Tri./Edg. & $\Delta t$ (sec) & $\#$ Iter. & Time (ms)\\
\whline{0.5pt}
Light & $16$ & 1.2K/796 & \begin{tabular}{@{}c@{}}1/100 \\ 1/240 \\ 1/1000\end{tabular}
& \begin{tabular}{@{}c@{}}1.9 \\ 1.5 \\ 1.1\end{tabular} 
& \begin{tabular}{@{}c@{}}\textbf{3} | \red{2} \\ \textbf{2.2} | \red{1.5} \\ \textbf{2} | \red{3} \end{tabular} \\
\whline{0.5pt}
Medium & 142 & 3.5K/2.3K & \begin{tabular}{@{}c@{}}1/100 \\ 1/240 \\ 1/1000\end{tabular}
& \begin{tabular}{@{}c@{}}7.6 \\ 2.9 \\ 1.3\end{tabular} 
& \begin{tabular}{@{}c@{}}\textbf{92} | \red{68} \\ \textbf{41} | \red{58} \\ \textbf{19} | \red{82} \end{tabular} \\
\whline{0.5pt}
Full & 562 & 11K/7.3K & \begin{tabular}{@{}c@{}} 1/100 \\ 1/240 \\ 1/1000\end{tabular}
& \begin{tabular}{@{}c@{}}11.0 \\ 4.4 \\ 1.8\end{tabular} 
& \begin{tabular}{@{}c@{}}\textbf{657} | \red{629} \\ \textbf{328} | \red{809} \\ \textbf{102} | \red{804} \end{tabular} \\
\whline{1.15pt}
\end{tabular}
\end{center}
}
\caption{\textbf{ABD vs. \texttt{Bullet} timing.}~~We record timing information of simulating the wrecking ball scenarios under different block counts and time step sizes using ABD and \texttt{Bullet}. \textbf{$\#$~Bdy} is the number of bodies in the test. \textbf{$\#$~Tri./Edg.} gives the total numbers of triangles and edges on the surface of the models. $\bm{\Delta t}$ is the time step size. \textbf{$\#$~Iter.} is the average iteration counts in ABD simulation. \textbf{Time} reports the average computation time for each frame using \textbf{ABD} and \red{\texttt{Bullet}}. This timing comparison is reported on a single thread implementation.}\label{tab:bullet}
\vspace{-20 pt}
\end{table}

\begin{figure*}[t!]
  \centering
  \includegraphics[width=\linewidth]{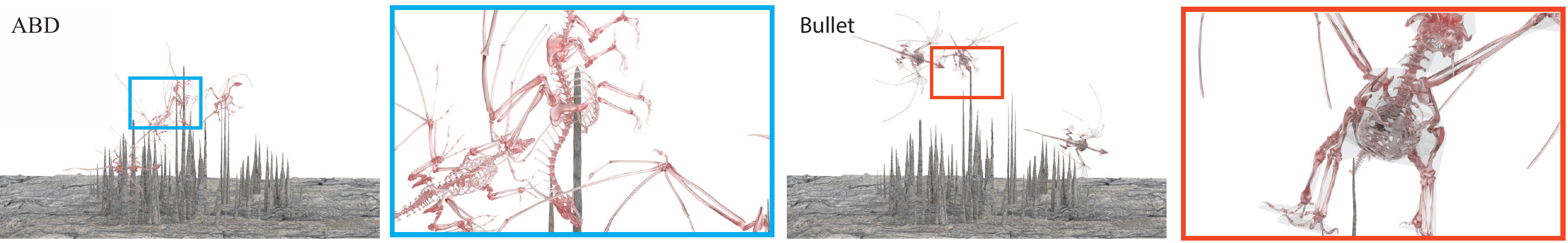}
  \caption{\textbf{Bone dragon.}~~In addition to the simulation algorithm, another difference between ABD and \texttt{Bullet} is the collision detection. ABD allows triangle-level collision detection to capture local contacts between fine geometries. \texttt{Bullet} is mesh-based and relies on convex decomposition of the model. In some cases, the decomposition is not accurate enough to approximate concave shapes. This figure shows one of such examples. Both bone dragons and rock spikes are sharp and concave.} \label{fig:bone_dragon}
\end{figure*}

The detailed timing comparison is listed in Tab.~\ref{tab:bullet}. In order to ensure the comparison is fair, we turn off multi-threading in ABD and \texttt{Bullet}. This is because our method processes collision on the surface triangles, which will receive more acceleration under parallelization. It is noted that under the light test ($\Delta t = 1/100$~sec), \texttt{Bullet} is rather efficient and only needs $2$~ms to simulate one frame. Our method is slower than \texttt{Bullet} and takes $3$~ms to simulate one frame. This difference diminishes under a smaller time step. For instance, when the time step size is set as $1/240$~sec, which is the default setting in \texttt{Bullet}, the difference of per-frame simulation time is less than one millisecond ($2.2$~ms for ABD and $1.5$~ms for \texttt{Bullet}). If the time step size is more conservatively set to $1/1000$~sec, ABD becomes faster than \texttt{Bullet} by a small margin ($2$~ms for ABD and $3$~ms for \texttt{Bullet}). However, \emph{simulations using \texttt{Bullet} in all time step sizes have intersections} (even with $\Delta t = 1/1000$~sec). Our method on the other hand is guaranteed to be free of inter-penetration. Another interesting observation is per-frame simulation using ABD consistently becomes more efficient under smaller time steps, which is not the case for \texttt{Bullet}. This may be because a smaller time step could expose more collisions to \texttt{Bullet} solver, which are otherwise missed under a bigger step. In addition, \texttt{Bullet} fails all the friction experiments including the screwing bolt, arch, and house of cards (Figs.~\ref{fig:friction}~and~\ref{fig:cards}) even with highly conservative time step size ($\Delta t = 1/10000$~sec).

The story is similar for the medium/full test: ABD ($92$~ms in medium and $657$~ms in full) is slightly slower than \texttt{Bullet} ($68$~ms in medium and $629$~ms in full) when $\Delta t = 1/100$~sec\footnote{Basically, ABD and \texttt{Bullet} have the same FPS in the full test with $\Delta t = 1/100$~sec}. Yet ABD takes the lead under smaller time steps of $1/240$~sec and $1/1000$~sec. In addition, \texttt{Bullet} has significantly more inter-penetration instances in the medium/full tests than in the light test, which could end up with undesired artifacts in many simulation tasks.

As discussed, ABD's collision processing directly handles each body's input surface mesh boundary with all guarantees, including non-intersection, applying directly to those meshes. This means that control of resolution and so quality is available via standard, well optimized geometry pipelines, e.g. mesh decimation. \texttt{Bullet}, however, requires pre-processing to convert all input meshes body into convex decompositions. In turn \texttt{Bullet} will then simulate these decomposed, simplified models as proxies, resolving collision handling on them, for each rigid body's surface rather than the original mesh. \texttt{Bullet} users often use V-HACD~\cite{mamou2016volumetric}, an automated library for computing convex decomposition, for this process. Utilizing libraries like V-HACD, can be a slow and often time-consuming iterative process to hand-tune quality parameters in order to obtain a reasonable a reasonable geometric approximation and resolution. Even so important features can be lost while details and symmetries are often unnecessarily broken. As such, many advanced users often resort to laborious hand-crafting of proxies when capturing important surface features is critical in an application (e.g., for design or robotics). Here, we demonstrate a comparative example of utilizing detailed, bone dragons in Fig~\ref{fig:bone_dragon} dropped on a highly featured geometry. After careful hand-tuning of V-HACD parameters, we present in Fig~\ref{fig:bone_dragon} right the resulting simulation using our best resulting V-HACD proxies. Here we see the convex decomposition geometry is still insufficiently detailed to capture local collision behavior between sharp asperities and convexities. In contrast, in Fig~\ref{fig:bone_dragon} left, ABD directly simulates the detailed geometry with tight tolerance so that all the original input meshes' detailed affordances are kept. In turn we see that this allows the dragons' concavities to tightly entangle and also slide directly onto and be caught by the sharp points -- all while remaining intersection free.

\subsection{Joint Constraints are Linear for Affine Bodies}\label{subsec:constraint}
An additional benefit of ABD is convenient constraint handling -- especially for the those prescribing rotational DOFs. For instance, it is common to require multiple objects to obey a given kinematic relation following the joint connecting them. Such constraints are often nonlinear as they are formulated from rigid body rotational DOF. Enforcing them would then require computing the derivatives of the rotational DOFs (either via constrained Lagrangian methods or generalized coordinates). Here relaxation from rigidity to affinity also eases the processing of such constraints.

It is known that a non-degenerate tetrahedron uniquely defines an affine transform. This suggests our generalized coordinate $q$ can be mapped to \emph{any} linear tetrahedron. From this perspective, an affine body simulation can also be viewed as a single-element FEM (with a simplified strain energy). The geometry of this element however, can be setup flexibly even its position deviates away significantly from the object. Let $\phi$ denote the map between $q$ and this virtual tetrahedron such that $q = \phi(\mathsf{P})$. $\mathsf{P}\in\mathbb{R}^{3 \times 4}$ stores the deformed vertex positions (i.e., $p_1$, $p_2$, $p_3$, and $p_4$) of the element. Let the rest shape position of the element be $\bar{\mathsf{P}}$, and one can easily verify that:
\begin{equation}\label{eq:phi}
\phi(\mathsf{P}) = \left[\frac{1}{4}\sum_i(p_i - \bar{p}_i)^T,\text{vec}^T\left(\mathsf{P}\bar{\mathsf{P}}^T(\bar{\mathsf{P}} \bar{\mathsf{P}}^T)^{-1}\right)\right]^T.
\end{equation}
Here, $\text{vec}(\cdot)$ denotes the vectorization of a matrix. Since $\bar{\mathsf{P}}$ is given, we could use $\text{vec}(\mathsf{P})$ as the new generalized coordinate of the system. Clearly $\phi$ is a linear function of $\mathsf{P}$. $\partial \phi/\partial p_i$ only depends on $\bar{\mathsf{P}}$. Therefore, the Jacobi of the system remains constant:
\begin{equation}\label{eq:tet_jacobi}
\mathsf{J}_i = \frac{\partial x_i}{\partial q} \cdot \frac{\partial \phi}{\partial \text{vec}(\mathsf{P})}\in\mathbb{R}^{3\times12}.
\end{equation}

\begin{figure}[ht!]
  \centering
  \includegraphics[width=\linewidth]{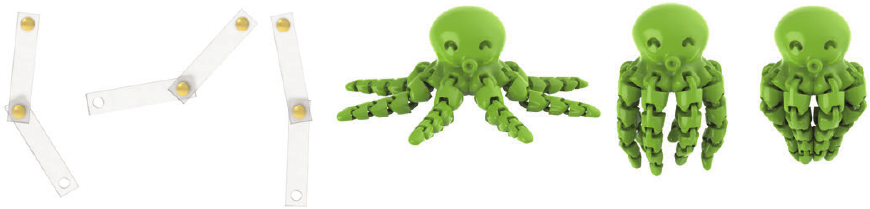}
  \caption{\textbf{Constrained simulation.}~~Prescribing rotation freedom in ABD is convenient. We visualize the generalized coordinate as a virtual tetrahedron, and most rotational constraints are linearized. We show two examples of such simulation in this figure. Both pendulum and octopus tentacles are made of multiple bodies connected via hinge joints. ABD remains significantly faster than rigid-IPC: ABD achieves a $371\times$ speedup for the pendulum and a $28\times$ speedup for the octopus.} \label{fig:rotation}
\end{figure}

\begin{figure*}[t!]
  \centering
  \includegraphics[width=\linewidth]{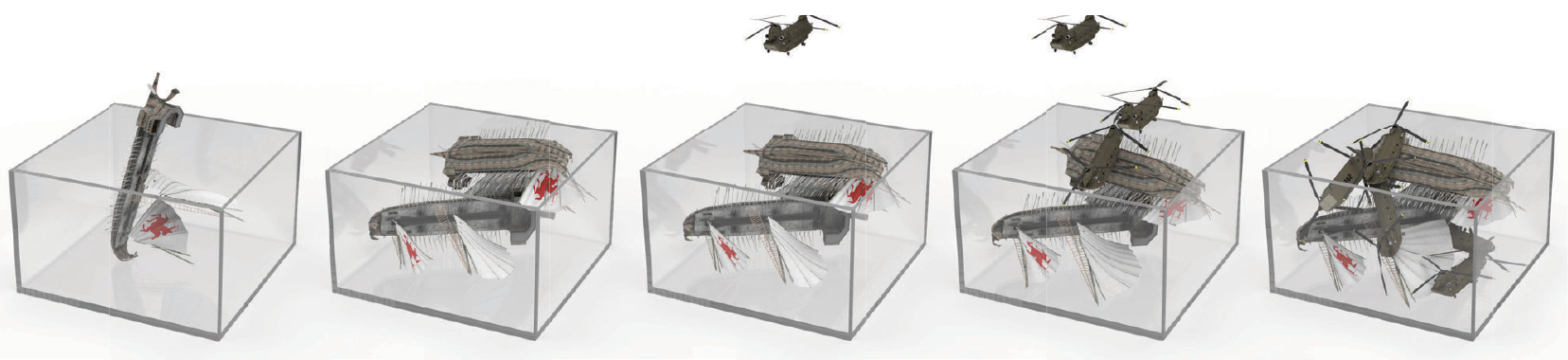}
  \caption{\textbf{Hybrid simulation.}~~ABD is particularly convenient for simulating hybrid object with both rigid and deformable parts. Both the barbarian ship ($225$K triangles, $341$K edges) and helicopter ($116$K triangles, $176$K edges) are such hybrid objects with a rigid main body appended by several deformable parts (canvas, rotors, and wheels). We use standard neo-Hookean FEM to simulate deformable parts, which are constrained to the virtual tetrahedron corresponding to the rigid body. Due to the involvement of massive DOFs from deformable parts, the simulation uses about $16$~sec for one frame under $\Delta t = 1/100$~sec.} \label{fig:hybrid}
\end{figure*}

\setlength{\columnsep}{10 pt}
\begin{wrapfigure}{r}{0.35\linewidth}
\includegraphics[width = \linewidth]{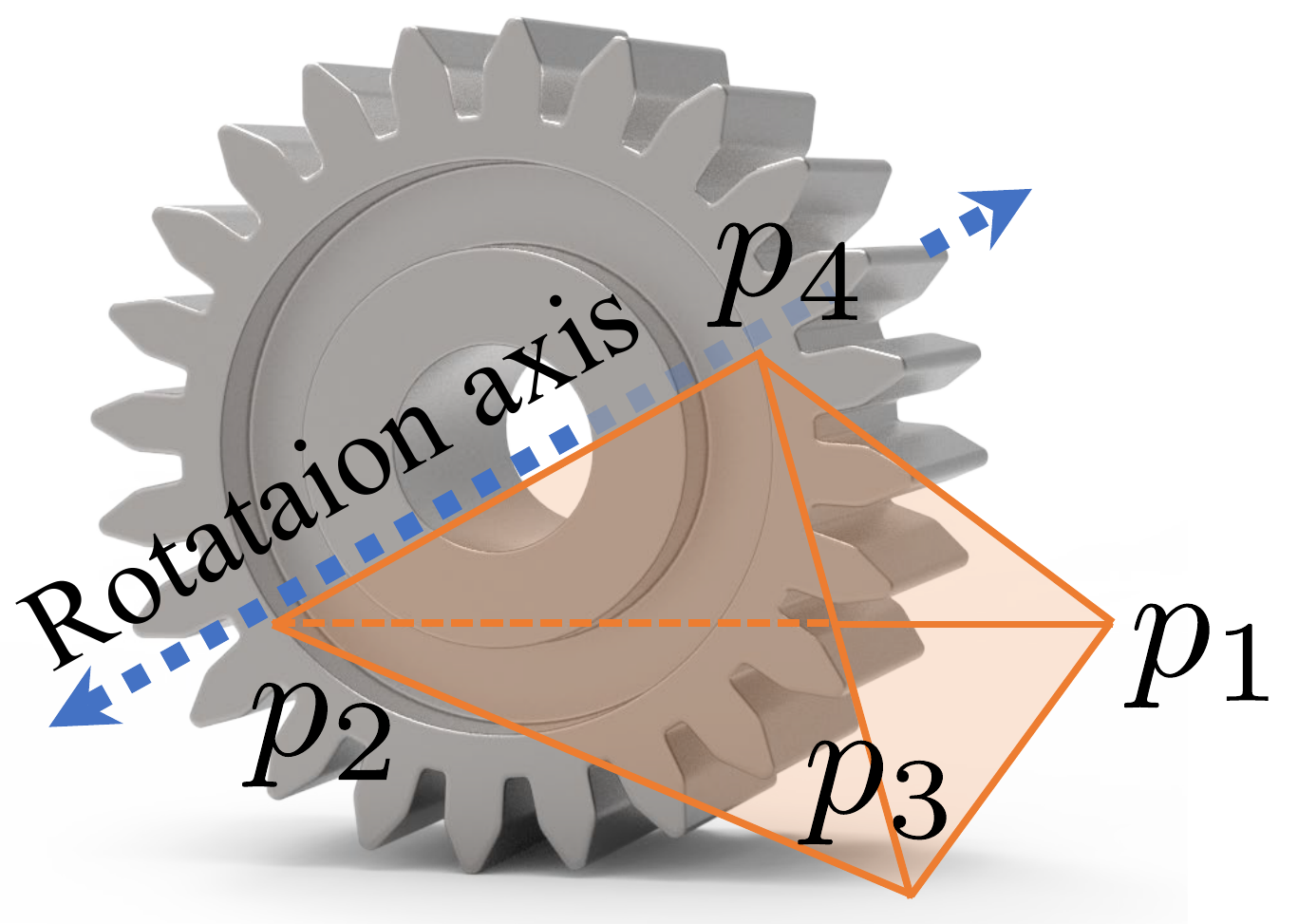}
\caption{\textbf{Convenient constraint handling.}~~An axis constraint becomes a linear one and can be easily enforced in A-IPC framework.}
\vspace{-5 pt}
\label{fig:tet}
\vspace{-5 pt}
\end{wrapfigure}
In practice, we exploit the fact that $\bar{\mathsf{P}}$ could be any tetrahedron to manipulate such constraints intuitively. For instance as shown in Fig.~\ref{fig:tet}, the gear is constrained to rotate around a prescribed axis (blue dash line). In ABD, we map $q$ to a tetrahedron. Since the tetrahedron can be chosen freely, we make sure that one of its edges is parallel to the rotation axis. After that, the axis constraint can be simply posed as a linear equality constraint fixing two vertices of the edge ($p_2$ and $p_4$ in the figure). Indeed, such formulation is over-constraining as it eliminates six freedoms instead of two (we should also allow $p_2$ and $p_4$ to move laterally along the edge). However, such ``over-constrainment'' still gives correct simulation because ABD uses more DOFs to simulate a rigid object. Other types constraints can be dealt with in a similar way (e.g., the constraint over the rotation angle can be converted to a linear shearing constraint).

\begin{figure}[h!]
  \centering
  \includegraphics[width=\linewidth]{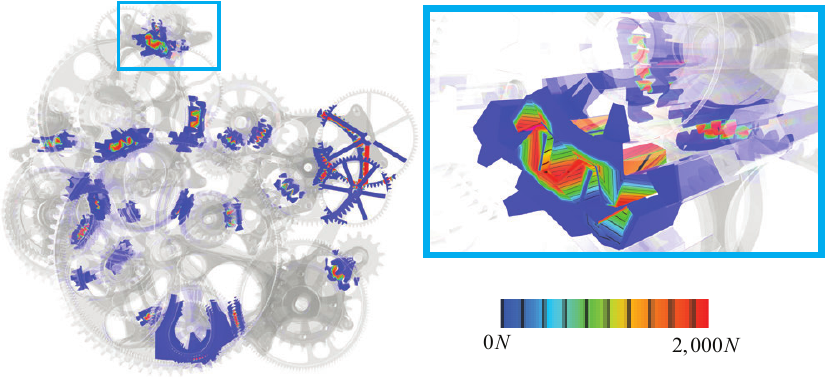}
  \caption{\textbf{Contact forces at teeth.}~~With the assistance of i-AABB, ABD efficiently and accurately model the contact forces among gear teeth. This figure visualizes the force distribution over the gear teeth.} \label{fig:gear}
\end{figure}

Two of such examples are reported in Fig.~\ref{fig:rotation}. The pendulum is a basic mechanism structure with two rigid links constrained by a hinge. The octopus has eight tentacles, and each of it consists of five joints. We also compared ABD with rigid-IPC with these models. ABD does not only enjoy a more handy formulation but also runs significantly faster ($371\times$ and $28\times$ speedups for the pendulum and octopus respectively). A more challenging experiment is reported in Fig.~\ref{fig:teaser}, where we simulate a more complex mechanism device of a set of 28 gears. Those gears are coupled by tooth contacts and shafts (see Fig.~\ref{fig:gear}). There are totally $2,450$K triangles and $3,090$K edges on the surface (i.e., those could participate in culling and CCD). A component-wise breakdown showing the inter-connectivity of gears is given in the right of the figure. As the driving torque is applied at one of the gear (the one with red arrow), the entire device moves forward. This simulation is very demanding for the robustness of the simulation algorithm. Because gear tooth are close to each other, highly localized and detailed CCD is massively used. All rigid body simulation algorithms we have tested fail in this case including rigid-IPC, \texttt{MuJuCo}, and \texttt{Bullet}. As mentioned, \texttt{Bullet} requires building a convex proxy for collision processing, which does not capture the zigzag geometry at the gear tooth. \texttt{MuJuCo} fails the test even under highly conservative settings (e.g., very small time step size). Rigid-IPC also fails this experiment: after first few steps, the Newton iteration loops infinitely because the cured CCD is unable to find a usable TOI (time of impact). ABD manages to simulate this scene without any issues. Each frame takes about $12$~sec on the CPU. As the majority of the computation is at CCD processing, we port our i-AABB algorithm to \texttt{CUDA}, and the simulation of the gear set reaches an interactive rate (at 5 -- 10 FPS).

\begin{figure*}[t!]
  \centering
  \includegraphics[width=\linewidth]{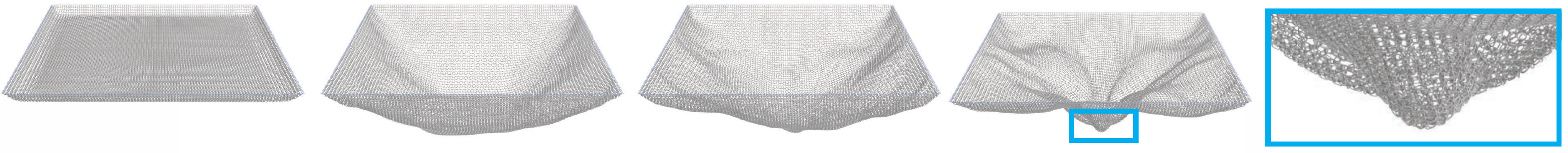}
  \caption{\textbf{Huge chain net.}~~When the body count increases, the disadvantage of using redundant DOFs in the simulation may become more obvious. Following this thought, we up scale the chain net to have $27,645$ bodies. In this stress test, A-IPC takes about $5$~min to simulate one frame on average. The entire simulation takes about five \emph{days} with A-IPC. We are never able to finish this test with rigid-IPC. Based on our observation, A-IPC is at least $2,000\times$ faster than rigid-IPC. This means it will need several \emph{years} for rigid-IPC to finish this experiment.} \label{fig:net_huge}
\end{figure*}

\subsection{Hybrid Simulation}
\label{subsec:hybrid}

Another benefit of ABD, when viewed as a reduced single-tetrahedron FEM, is to simulate models with both rigid and soft components. Such objects are vastly available in real world. While existing rigid body methods can also be augmented to incorporate such hybrid simulations, it is particularly effortless in the framework of ABD. We show a simulation of such scene. The barbarian ship has a rigid ship body and two deformable canvas. Each ship has $225$K surface triangles and $341$K edges. The helicopter is also hybrid with a rigid body, two soft rotors, and four soft wheels. There are $116$K triangles and $176$K edges on each helicopter. After two ships fall into the glass tank, five helicopters follow, producing interesting animation effects of both rigid and deformable dynamics. Note that the collision between rigid and soft bodies can be handled uniformly using barrier-based penalties. In such hybrid simulation, the Hessian of elastic potential on the deformable components are much smaller (by several orders) than the Hessian of the orthogonality potential ($V_{\perp}$), and the positive definiteness of the global system matrix is numerically compromised. Therefore, LLT factorization (\texttt{SimplicialLLT}) may fail occasionally. In this case, we switch to LDLT Cholesky for each Newton solve. A possible remedy for increase numerically stability is to use Schur complement like formulation~\cite{peiret2019schur} to somehow decouple DOFs from rigid (affine) and deformable components.

\begin{table*}[th!]
\caption{\textbf{Time statistic.}~~Detailed time statistics of our experiments. In most experiments, we tested ABD under three different $\Delta t$ settings namely $1/100$, $1/50$, and $1/25$. \textbf{$\#$~Bdy} (number of bodies), \textbf{$\#$~Tri./Edg.} (numbers of triangles and edges), $\bm{\Delta t}$ (time step size), \textbf{$\#$~Iter.} (average per-frame iteration counts), and \textbf{Time} (total time for each frame) are the same measures as in Tab.~\ref{tab:bullet}. This table also reports computation time used for Hessian assembly (\textbf{Hess.}), total time used for solving the linear system in Newton method at each frame (\textbf{Sol.}), time used for triangle-level CCD (\textbf{CCD}), time used for building the collision pairs (\textbf{Cons.}), and other computations (\textbf{Misc.} e.g., variables initialization, convergence check etc. In many examples (except for the last four examples), we give comparative timing information of both \textbf{ABD} and \teal{rigid-IPC}. The \textbf{\blue{GPU FPS}} is also reported for $\Delta t = 1/100$ ABD simulations.}\label{tab:time}
{\footnotesize \fontfamily{ppl}\selectfont
\begin{center}
\begin{tabular}{c|c|c|c|c|c|c|c|c|c|c}
\whline{1.15pt}
Test & $\#$ Bdy & $\#$ Tri./Edg. & $\Delta t$ (sec) & $\#$ Iter. & Hess. (sec) & Sol. (ms) & CCD (sec) & Cons. (sec) &  Misc. (ms) & Time (sec)\\
\whline{0.5pt}
\begin{tabular}{@{}c@{}} Wrecking ball \\ (Fig.~\ref{fig:wrecking})  \end{tabular}
& 575 & 14K/20K & \begin{tabular}{@{}c@{}}1/100 \\ 1/50 \\ 1/25\end{tabular}
& \begin{tabular}{@{}c@{}} \textbf{8.8} | \teal{17.1} \\ \textbf{19.4} \\ \textbf{42.6} \end{tabular} 
& \begin{tabular}{@{}c@{}} \textbf{0.023} | \teal{10.3} \\ \textbf{0.067} \\ \textbf{0.18} \end{tabular} 
& \begin{tabular}{@{}c@{}}\textbf{53} | \teal{47} \\ \textbf{160}  \\ \textbf{444} \end{tabular} 
& \begin{tabular}{@{}c@{}}\textbf{0.028} | \teal{4.9} \\ \textbf{0.078}  \\ \textbf{0.23} \end{tabular}
& \begin{tabular}{@{}c@{}}\textbf{0.031} | \teal{2.3} \\ \textbf{0.092}  \\ \textbf{0.027} \end{tabular}
& \begin{tabular}{@{}c@{}}\textbf{8} | \teal{53} \\ \textbf{18}  \\ \textbf{43} \end{tabular}
& \begin{tabular}{@{}c@{}}\textbf{0.14} | \teal{17.6} | \blue{\textbf{18}{\tiny{ FPS}}}\\ \textbf{0.41}  \\ \textbf{1.1} \end{tabular}\\
\whline{0.5pt}
\begin{tabular}{@{}c@{}} Small chain net \\ (Fig.~\ref{fig:net})  \end{tabular}
& 144 & 63K/95K & \begin{tabular}{@{}c@{}}1/100 \\ 1/50 \\ 1/25\end{tabular}
& \begin{tabular}{@{}c@{}} \textbf{2.1} | \teal{4.0} \\ \textbf{2.4} \\ \textbf{2.8} \end{tabular} 
& \begin{tabular}{@{}c@{}} \textbf{0.011} | \teal{0.18} \\ \textbf{0.014} \\ \textbf{0.018} \end{tabular} 
& \begin{tabular}{@{}c@{}}\textbf{6} | \teal{2} \\ \textbf{7}  \\ \textbf{9} \end{tabular} 
& \begin{tabular}{@{}c@{}}\textbf{0.016} | \teal{1.1} \\ \textbf{0.023}  \\ \textbf{0.29} \end{tabular}
& \begin{tabular}{@{}c@{}}\textbf{0.017} | \teal{0.73} \\ \textbf{0.023}  \\ \textbf{0.032} \end{tabular}
& \begin{tabular}{@{}c@{}}\textbf{9} | \teal{6} \\ \textbf{10}  \\ \textbf{10} \end{tabular}
& \begin{tabular}{@{}c@{}}\textbf{0.059} | \teal{2.1} | \blue{\textbf{143}{\tiny{ FPS}}}\\ \textbf{0.07}  \\ \textbf{0.08} \end{tabular}\\
\whline{0.5 pt}
\begin{tabular}{@{}c@{}} Big chain net \\ (Fig.~\ref{fig:net_big})  \end{tabular}
& 673 & 445K/297K & \begin{tabular}{@{}c@{}}1/100 \\ 1/50 \\ 1/25\end{tabular}
& \begin{tabular}{@{}c@{}} \textbf{13.5} | \teal{169} \\ \textbf{24.9} \\ \textbf{26.3} \end{tabular} 
& \begin{tabular}{@{}c@{}} \textbf{0.13} | \teal{109} \\ \textbf{0.25} \\ \textbf{0.27} \end{tabular} 
& \begin{tabular}{@{}c@{}}\textbf{241} | \teal{437} \\ \textbf{440}  \\ \textbf{464} \end{tabular} 
& \begin{tabular}{@{}c@{}}\textbf{0.41} | \teal{305} \\ \textbf{0.82}  \\ \textbf{0.93} \end{tabular}
& \begin{tabular}{@{}c@{}}\textbf{0.43} | \teal{388} \\ \textbf{0.87}  \\ \textbf{0.96} \end{tabular}
& \begin{tabular}{@{}c@{}}\textbf{62} | \teal{1336} \\ \textbf{83}  \\ \textbf{86} \end{tabular}
& \begin{tabular}{@{}c@{}}\textbf{0.78} | \teal{944} | \blue{\textbf{5}{\tiny{ FPS}}} \\ \textbf{2.4}  \\ \textbf{2.7} \end{tabular}\\
\whline{0.5 pt}
\begin{tabular}{@{}c@{}} House of cards \\ (Fig.~\ref{fig:cards})  \end{tabular}
& 158 & 336/816 & \begin{tabular}{@{}c@{}}1/100 \\ 1/50 \\ 1/25\end{tabular}
& \begin{tabular}{@{}c@{}} \textbf{9.8} | \teal{66.5} \\ \textbf{13.3} \\ \textbf{25.2} \end{tabular} 
& \begin{tabular}{@{}c@{}} \textbf{0.014} | \teal{2.49} \\ \textbf{0.024} \\ \textbf{0.045} \end{tabular} 
& \begin{tabular}{@{}c@{}}\textbf{24} | \teal{41} \\ \textbf{37}  \\ \textbf{64} \end{tabular} 
& \begin{tabular}{@{}c@{}}\textbf{0.009} | \teal{5.1} \\ \textbf{0.014}  \\ \textbf{0.035} \end{tabular}
& \begin{tabular}{@{}c@{}}\textbf{0.035} | \teal{1.1} \\ \textbf{0.043}  \\ \textbf{0.074} \end{tabular}
& \begin{tabular}{@{}c@{}}\textbf{4} | \teal{46} \\ \textbf{20}  \\ \textbf{9} \end{tabular}
& \begin{tabular}{@{}c@{}}\textbf{0.086} | \teal{8.9} | \blue{\textbf{42}{\tiny{ FPS}}} \\ \textbf{0.13}  \\ \textbf{0.24} \end{tabular}\\
\whline{0.5 pt}
\begin{tabular}{@{}c@{}} Screw \\ (Fig.~\ref{fig:friction}, left)  \end{tabular}
& 2 & 7.9K/5.2K & \begin{tabular}{@{}c@{}}1/100 \\ 1/50 \\ 1/25\end{tabular}
& \begin{tabular}{@{}c@{}} \textbf{4.1} | \teal{8.2} \\ \textbf{3.2} \\ \textbf{5.7} \end{tabular} 
& \begin{tabular}{@{}c@{}} \textbf{0.0004} | \teal{0.028} \\ \textbf{0.0007} \\ \textbf{0.001} \end{tabular} 
& \begin{tabular}{@{}c@{}}\textbf{0.1} | \teal{0.1} \\ \textbf{0.1}  \\ \textbf{0.1} \end{tabular} 
& \begin{tabular}{@{}c@{}}\textbf{0.039} | \teal{2.5} \\ \textbf{0.068}  \\ \textbf{0.11} \end{tabular}
& \begin{tabular}{@{}c@{}}\textbf{0.039} | \teal{0.096} \\ \textbf{0.068}  \\ \textbf{0.12} \end{tabular}
& \begin{tabular}{@{}c@{}}\textbf{9} | \teal{1} \\ \textbf{10}  \\ \textbf{10} \end{tabular}
& \begin{tabular}{@{}c@{}}\textbf{0.087} | \teal{2.6} | \blue{\textbf{1K}{\tiny{ FPS}}} \\ \textbf{0.14}  \\ \textbf{0.23} \end{tabular}\\
\whline{0.5 pt}
\begin{tabular}{@{}c@{}} Arch \\ (Fig.~\ref{fig:friction}, right)  \end{tabular}
& 101 & 1.2K/1.8K & \begin{tabular}{@{}c@{}}1/100 \\ 1/50 \\ 1/25\end{tabular}
& \begin{tabular}{@{}c@{}} \textbf{1.9} | \teal{6.2} \\ \textbf{3.2} \\ \textbf{5.7} \end{tabular} 
& \begin{tabular}{@{}c@{}} \textbf{0.002} | \teal{0.13} \\ \textbf{0.003} \\ \textbf{0.005} \end{tabular} 
& \begin{tabular}{@{}c@{}}\textbf{2} | \teal{44} \\ \textbf{2}  \\ \textbf{3} \end{tabular} 
& \begin{tabular}{@{}c@{}}\textbf{0.001} | \teal{0.44} \\ \textbf{0.001}  \\ \textbf{0.003} \end{tabular}
& \begin{tabular}{@{}c@{}}\textbf{0.002} | \teal{0.099} \\ \textbf{0.003}  \\ \textbf{0.006} \end{tabular}
& \begin{tabular}{@{}c@{}}\textbf{2} | \teal{3} \\ \textbf{2}  \\ \textbf{3} \end{tabular}
& \begin{tabular}{@{}c@{}}\textbf{0.0087} | \teal{0.67} | \blue{\textbf{500}{\tiny{ FPS}}}\\ \textbf{0.013}  \\ \textbf{0.022} \end{tabular}\\
\whline{0.5 pt}
\begin{tabular}{@{}c@{}} Pendulum \\ (Fig.~\ref{fig:rotation}, left)  \end{tabular}
& 4 & 1.3K/2.0K & \begin{tabular}{@{}c@{}}1/100 \\ 1/50 \\ 1/25\end{tabular}
& \begin{tabular}{@{}c@{}} \textbf{4.1} | \teal{4.3} \\ \textbf{4.3} \\ \textbf{4.7} \end{tabular} 
& \begin{tabular}{@{}c@{}} \textbf{0.001} | \teal{0.018} \\ \textbf{0.001} \\ \textbf{0.001} \end{tabular} 
& \begin{tabular}{@{}c@{}}\textbf{0.1} | \teal{0.1} \\ \textbf{0.1}  \\ \textbf{0.1} \end{tabular} 
& \begin{tabular}{@{}c@{}}\textbf{0.003} | \teal{2.5} \\ \textbf{0.004}  \\ \textbf{0.005} \end{tabular}
& \begin{tabular}{@{}c@{}}\textbf{0.003} | \teal{0.03} \\ \textbf{0.003}  \\ \textbf{0.004} \end{tabular}
& \begin{tabular}{@{}c@{}}\textbf{1} | \teal{1} \\ \textbf{1}  \\ \textbf{1} \end{tabular}
& \begin{tabular}{@{}c@{}}\textbf{0.007} | \teal{2.6} | \blue{\textbf{1K}{\tiny{ FPS}}}\\ \textbf{0.009}  \\ \textbf{0.01} \end{tabular}\\
\whline{0.5 pt}
\begin{tabular}{@{}c@{}} Octopus \\ (Fig.~\ref{fig:rotation}, right)  \end{tabular}
& 41 & 33K/49K & \begin{tabular}{@{}c@{}}1/100 \\ 1/50 \\ 1/25\end{tabular}
& \begin{tabular}{@{}c@{}} \textbf{4.6} | \teal{4.3} \\ \textbf{5.6} \\ \textbf{6.7} \end{tabular} 
& \begin{tabular}{@{}c@{}} \textbf{0.012} | \teal{0.071} \\ \textbf{0.014} \\ \textbf{0.017} \end{tabular} 
& \begin{tabular}{@{}c@{}}\textbf{2} | \teal{0.5} \\ \textbf{0.2}  \\ \textbf{0.2} \end{tabular} 
& \begin{tabular}{@{}c@{}}\textbf{0.02} | \teal{0.97} \\ \textbf{0.03}  \\ \textbf{0.04} \end{tabular}
& \begin{tabular}{@{}c@{}}\textbf{0.019} | \teal{0.34} \\ \textbf{0.025}  \\ \textbf{0.03} \end{tabular}
& \begin{tabular}{@{}c@{}}\textbf{5} | \teal{3} \\ \textbf{6}  \\ \textbf{6} \end{tabular}
& \begin{tabular}{@{}c@{}}\textbf{0.05} | \teal{1.4} | \blue{\textbf{333}{\tiny{ FPS}}}\\ \textbf{0.07}  \\ \textbf{0.09} \end{tabular}\\
\whline{0.5 pt}
\begin{tabular}{@{}c@{}} Hybrid sim. \\ (Fig.~\ref{fig:hybrid})  \end{tabular}
& 19+8 & 1.1M/1.6M & \begin{tabular}{@{}c@{}}1/100 \\ 1/50 \\ 1/25\end{tabular}
& \begin{tabular}{@{}c@{}} \textbf{16.7} | \teal{--} \\ \textbf{19.3} \\ \textbf{25.9} \end{tabular} 
& \begin{tabular}{@{}c@{}} \textbf{4.3} | \teal{--} \\ \textbf{5.8} \\ \textbf{7.9} \end{tabular} 
& \begin{tabular}{@{}c@{}}\textbf{4600} | \teal{--} \\ \textbf{5311}  \\ \textbf{7083} \end{tabular} 
& \begin{tabular}{@{}c@{}}\textbf{3.5} | \teal{--} \\ \textbf{4.8}  \\ \textbf{7.2} \end{tabular}
& \begin{tabular}{@{}c@{}}\textbf{3.4} | \teal{--} \\ \textbf{4.5}  \\ \textbf{6.9} \end{tabular}
& \begin{tabular}{@{}c@{}}\textbf{490} | \teal{--} \\ \textbf{562}  \\ \textbf{836} \end{tabular}
& \begin{tabular}{@{}c@{}}\textbf{16.4} | \teal{--} | \blue{\textbf{0.4}{\tiny{ FPS}}} \\ \textbf{21.3}  \\ \textbf{30.0} \end{tabular}\\
\whline{0.5 pt}
\begin{tabular}{@{}c@{}} Gear set \\ (Fig.~\ref{fig:teaser})  \end{tabular}
& 28 & 2.5M/3.1M & \begin{tabular}{@{}c@{}}1/100 \\ 1/50 \\ 1/25\end{tabular}
& \begin{tabular}{@{}c@{}} \textbf{16.7} | \teal{--} \\ \textbf{19.3} \\ \textbf{25.9} \end{tabular} 
& \begin{tabular}{@{}c@{}} \textbf{4.3} | \teal{--} \\ \textbf{5.8} \\ \textbf{7.9} \end{tabular} 
& \begin{tabular}{@{}c@{}}\textbf{4} | \teal{--} \\ \textbf{4}  \\ \textbf{6} \end{tabular} 
& \begin{tabular}{@{}c@{}}\textbf{5.4} | \teal{--} \\ \textbf{6.3}  \\ \textbf{9.3} \end{tabular}
& \begin{tabular}{@{}c@{}}\textbf{6.0} | \teal{--} \\ \textbf{7.9}  \\ \textbf{11.2} \end{tabular}
& \begin{tabular}{@{}c@{}}\textbf{211} | \teal{--} \\ \textbf{219}  \\ \textbf{43} \end{tabular}
& \begin{tabular}{@{}c@{}}\textbf{11.7} | \teal{--} | \blue{\textbf{7.3}{\tiny{ FPS}}} \\ \textbf{14.5}  \\ \textbf{20.1} \end{tabular}\\
\whline{0.5 pt}
\begin{tabular}{@{}c@{}} Bone dragon \\ (Fig.~\ref{fig:bone_dragon})  \end{tabular}
& 29 & 1.2M/1.7M 
& 1/100 
& \textbf{6.3} 
& \textbf{0.003} 
& \textbf{0.2}
& \textbf{2.3}
& \textbf{0.56}
& \textbf{74}
& \textbf{2.8} \\
\whline{0.5 pt}
\begin{tabular}{@{}c@{}} Huge chain net \\ (Fig.~\ref{fig:net_huge})  \end{tabular}
& 27,645 & 12M/18M 
& 1/100 
& \textbf{14.2} 
& \textbf{37.2} 
& \textbf{87~sec}
& \textbf{95}
& \textbf{94}
& \textbf{7.2~sec}
& \textbf{310} \\
\whline{1.15pt}
\end{tabular}
\end{center}
}
\end{table*}

\subsection{Timing and Breakdown}\label{subsec:timing}
Detailed time statistics of other the experiments is reported in Tab.~\ref{tab:time}. In all the experiments, we uniformly scale the scene to a $1 \times 1 \times 1$ box and set $\hat{d}$ as $1/1000$. In other words, if the size of the model is around one meter, the contact accuracy is guaranteed to be less than one millimeter, and all the models do not intersect with each other at any time during the simulation. We report comparative timing benchmark of ABD and rigid-IPC under $\Delta t = 1/100$ in most experiments. Normally, rigid-IPC is numerically robust under larger time steps. However, its performance is highly sensitive to a bigger $\Delta t$. This is because the underlying curved CCD quickly becomes prohibitive when examining over a wider trajectory gap. Therefore, we do not report our speedup over rigid-IPC for any time steps bigger than $\Delta t = 1/100$~sec. For chain net example shown in Fig.~\ref{fig:net_big}, the literal speedup exceeds four orders in general if we set $\Delta t = 1/25$. It is not our intention to oversell just by tweaking the time step size.

The majority portion of our performance improvement is brought by relaxing the rigidity constraint. This can be reflected by two important metrics from Tab.~\ref{tab:time}: the iteration count (\textbf{$\#$ Iter.}) and the time needed for CCD processing (\textbf{CCD}). It is clearly shown that as long as the contact frequency in the simulation is intense, ABD always requires much fewer iterations to converge than rigid-IPC does. This difference is ``extremized''  in the gear set example (Fig.~\ref{fig:teaser}), where ABD needs $17$ iterations for each time step, and rigid-IPC needs infinite iterations. CCD processing is another major game changer. As mentioned, rigid-IPC strictly follows the rigidity constraint making per-step trajectory curved. This curved trajectory is split and converted back to piece-wise line segments again during the CCD. This conversion is fully avoided in ABD, and one can use any existing CCD algorithm to detect potential intersection between primitives and compute TOI. We would like to remind that CCD needs to be carried out at each Newton iteration in order to make sure the line search does not introduce inter-penetrations. Therefore, the performance gap between A-IPC and rigid-IPC is further scaled by the iteration count. Together, those two factors contribute to $90\%$ of our speedup in complex simulations such as the big chain net (Fig.~\ref{fig:net_big}), house of cards (Fig.~\ref{fig:cards}) and wrecking ball (Fig.~\ref{fig:wrecking}). In other relatively lighter simulation experiments, our improved culling and Hessian assembly become more profitable. Our culling is up to to $90\%$ more effective than conventional BVH-based strategy for models with complex geometry. It is on average $30\%$ more effective for simpler shapes (e.g., the chain net or the octopus). In terms of system solve, ABD is typically slower than existing rigid body methods due to inflated DOFs. However, this disadvantage is invisible because ABD always enjoys a much fewer iterations and much faster CCD processing. This is our core inspiration of designing ABD.

We suspect increasing the body count will grant the lead to rigid-IPC at a certain point and simulate a huge chain net model with $27,645$ bodies. The result is opposite -- we estimate ABD has a $\sim5,000\times$ speedup (on CPU). This is just a rough assessment as we are never able to finish rigid-IPC in this stress test, which will need \emph{several years} while ABD finishes the simulation in \emph{five days}.

\section{Conclusion}
We have introduced a new, simple affine dynamics model and a carefully customized, easy-to-implement affine IPC algorithm for the simulation of extremely stiff materials with fidelity, convergence and reliability. The resulting method is highly suited for simulating all scenarios and applications where currently rigid body methods are now popularly employed \emph{without}, as we have shown, the current limitations that rigid body models impose. Here we have demonstrated that ABD obtains orders of magnitude speedup over state-of-the-art rigid body simulation with comparable guarantees of non-intersection and convergence. At the same time we have also shown that ABD obtains both comparable (for easy examples) and improved (as scene complexity grows) speeds when compared with highly optimized rigid body libraries that do not have guarantees and so suffer from artifacts and all-out failures that limit their automated use.

ABD is custom-suited for parallelization, is also differentiable, and automatically and directly simulates all input triangulated geometries. We have shown that, when leveraging the GPU, ABD can simulate complex contacting systems at interactive rates. With these combined properties it is then exciting to consider future applications where ABD's automation, reliability, and differentiability can be utilized for computational design, machine learning, and robotics. In these cases consistent, artifact-free simulation behavior across shape, material and contact variations, without algorithm parameter tuning, should accelerate development. We have also shown a few initial, proofs-of-concept for extensions of ABD to both complex, jointed stiff multibody systems, and to hybrid stiff/flexible multibody systems. Here there also clearly remain significant opportunities for further development and application. Finally, looking ahead, with the popularity and diverse applications of physical modeling we hope that ABD will provide the rapidly growing and diverse community of simulation users with a reliable, differentiable and exceedingly efficient framework suitable to swap in for all rigid body-type applications.

\bibliographystyle{ACM-Reference-Format}
\bibliography{aipc}
\end{document}